\documentclass[12pt]{article}
\usepackage{amsmath}
\usepackage{graphicx}
\usepackage{enumerate}
\usepackage{natbib}
\usepackage{url} 
\usepackage{bm}
\usepackage{subfigure}
\usepackage{xcolor}

\newcommand{\blind}{0}

\newcommand{\OHC}{\mathrm{OHC}}
\newcommand{\W}{\mathcal{W}}

\newcommand{\s}{\boldsymbol{s}}
\newcommand{\iid}{\overset{\text{i.i.d.}}{\sim}}
\newcommand{\Var}{\text{Var}}

\newcommand{\T}{\text{T}}
\newcommand{\up}{\text{up}}
\newcommand{\mi}{\text{mid}}
\newcommand{\lat}{\text{lat}}
\newcommand{\lon}{\text{lon}}
\renewcommand{\u}{\bm{u}}

\begin{document}

\def\spacingset#1{\renewcommand{\baselinestretch}%
{#1}\small\normalsize} \spacingset{1}


\if0\blind
{
  \title{\bf Improved Global Ocean Heat Content Estimation by Modeling Vertical Spatio-Temporal Dependence}
  \author{Thea Sukianto\hspace{.2cm}\\
    Department of Statistics and Data Science, Carnegie Mellon University\\
    Donata Giglio \\
    Department of Atmospheric and Oceanic Sciences,\\University of Colorado Boulder\\
    Mikael Kuusela \\
    Department of Statistics and Data Science, Carnegie Mellon University\\
    }
    \date{}
  \maketitle
} \fi

\if1\blind
{
  \bigskip
  \bigskip
  \bigskip
  \begin{center}
    {\LARGE\bf Improved Global Ocean Heat Content Estimation by Modeling Vertical Spatio-Temporal Dependence}
\end{center}
  \medskip
} \fi

\bigskip
\begin{abstract}
Estimating ocean heat content (OHC) with reliable uncertainties is critical for understanding and monitoring the evolution of Earth's climate, as the ocean has stored most of the energy accumulated in the climate system due to Earth Energy Imbalance. Here, we use Argo profiling float data from 2004-2022 to map OHC. As fewer Argo observations are available deeper in the water column, previous studies have partitioned the ocean into at least two pressure layers and mapped each separately, which complicates the estimation of uncertainties when the maps are summed to get the total OHC. In this work, we consider the case of two pressure layers and propose an improved mapping and uncertainty quantification method using bivariate locally stationary Gaussian processes and conditional simulations to map the two sections jointly while accounting for the correlation between them. We find that modeling this correlation results in improved OHC anomaly mapping and up to a 15\% reduction of global OHC anomaly uncertainties in comparison to mapping the two layers separately without accounting for their dependence. These estimated uncertainties are essential to analyze the statistical significance of OHC anomalies on both regional and global scales, which we demonstrate using several climatological case studies.
\end{abstract}

\noindent%
\vfill

\newpage
\spacingset{1.9} 
\section{Introduction}
\label{sec:intro}
The Earth Energy Imbalance (EEI) is the imbalance at the top of the atmosphere between the incoming solar radiation and the total outgoing radiation  \citep[i.e. Earth emitted energy and solar reflected radiation;][]{10.3389/fmars.2019.00432}. Estimating EEI is essential for understanding and monitoring the evolution of the Earth’s climate, yet it is difficult, as EEI is a globally integrated quantity whose magnitude and variations are small (of the order of 1 Wm$^{–2}$, \cite{Von_Schuckmann2016-rf}) compared to the amount of energy entering and leaving the Earth system ($\sim340$ Wm$^{–2}$ for solar irradiance, \cite{IPCC6}). Taking an inventory of the energy stored in different climate system reservoirs (the atmosphere, the land, the cryosphere and the ocean) and estimating their changes with time provides an indirect estimate of EEI \citep{10.3389/fmars.2019.00432}. Notably, estimating ocean heat content (OHC) is critical to estimate EEI, as the vast majority $(>90\%)$ of the energy uptake associated with EEI has occurred in the ocean (\cite{InsightsintoEarthsEnergyImbalancefromMultipleSources}). Producing OHC estimates with reliable uncertainties is all the more important as recent  intercomparison efforts (e.g., \cite{Hakuba2024}, \cite{Von_Schuckmann2023}) have showed a wide spread in OHC estimates based on varying data analysis pipelines and modeling assumptions. Studying changes in OHC is also key to understanding other aspects of how the Earth system is evolving, including sea level rise \citep{Johnson2022-sv} and changes in tropical cyclone intensity in a warming climate (e.g., \cite{Trenberth2018-jw}). 

Since the mid-2000s, the Argo program (e.g., \cite{ArgoDataJan2023}, \cite{Riser2016}) has provided  in-situ observations of ocean temperature and salinity (T/S)  on a nearly global scale, which are key to estimate OHC. Most of the floats in the Argo array descend to approximately 2000 dbar (1 dbar $\approx$ 1 m) and collect T/S profiles roughly every 10 days as they ascend the water column to the surface. As of this work, there are $\sim4100$ active floats that observe the global ocean at a $3^\circ \times 3^\circ$ nominal resolution.

Since OHC at a given time is an integrated quantity over latitude, longitude, and pressure, the Argo temperature profiles need to first be mapped onto a regular gridded field. However, the large volume and nonstationarity of the observations pose challenges to classical spatial interpolation techniques. As more than 3 million vertical profiles have been collected since the early 2000s, producing parameter estimates and point predictions using traditional spatio-temporal methods is computationally infeasible. Moreover, defining a nonstationary covariance function over the global ocean is not only computationally prohibitive, but is also unlikely to sufficiently capture regional-scale variability and dependence structure. In addition, a number of  profiles do not reach 2000 dbar due to instrument limitations or bathymetry (underwater land features such as continental shelves and seamounts). Accounting for these complexities in the oceanographic fields and floats' vertical sampling poses a delicate undertaking, especially in regard to the uncertainties.

In this paper, we introduce an extension of the univariate OHC mapping and uncertainty quantification framework developed in \cite{univariate} that incorporates the vertical dependence between two pressure layers. 
In particular, we use locally stationary bivariate Gaussian process regression to jointly model the upper and midocean OHC anomalies. Since the midocean layer has fewer observations, this improves the mapped anomaly fields by allowing the midocean to borrow predictive strength from the upper ocean. We also present a bivariate extension of the local conditional simulation algorithm in \cite{univariate}. This approach enables statistically rigorous, computationally tractable uncertainty estimates for global and regional OHC, which we now improve from the univariate approach's conservative bound. 

We compare the effect of incorporating vertical dependence with the previous univariate approach and find substantial benefits in a number of areas. Using cross-validation, we find that the bivariate model has improved leave-out predictive performance. This supports our observation that the predictive variance for the mapped anomalies is reduced by 10\%+, most prominently in the midocean section. 

Furthermore, we find that modeling the vertical dependence reduces uncertainties on downstream quantity estimates by as much as $15\%$. We demonstrate this uncertainty reduction on three key climatological estimates using the Argo data from 2004--2022: (1) the total OHC (upper + midocean) trend, (2) the total ocean heat uptake and its comparison with the top-of-atmosphere radiative flux, and (3) the cross-correlation between the total OHC anomaly and the Oceanic Nino Index. In particular, a rigorous uncertainty on the latter estimate would not have been possible to obtain without incorporating vertical dependence into the local conditional simulation realizations.

While we demonstrate our improved framework by jointly mapping OHC anomaly fields in two pressure layers, this work will facilitate further oceanographic applications with rigorously quantified uncertainties such as joint mapping of OHC and sea surface temperature (SST) or sea surface height (SSH) \citep{LymanJohnson23}. To encourage this goal, we provide a modular and reproducible codebase for our mapping method under a permissive open-source license at \url{https://github.com/ttsukianto/LocalGP_OHC}.

The rest of this paper is organized as follows: In Section \ref{sec:background}, we provide an overview of the Argo dataset, OHC estimation problem, and existing mapped products. In Section \ref{sec:methods}, we describe our previous univariate approach and introduce our improved OHC mapping and uncertainty quantification methods based on bivariate locally stationary Gaussian process regression and local conditional simulation; in Section \ref{sec:validation}, we validate the effect of modeling vertical dependence on the OHC estimates and uncertainties; in Section \ref{sec:results}, we demonstrate the downstream uncertainty improvements in a number of climatological case studies; finally, in Section \ref{sec:discussion}, we provide a discussion of further potential modeling and computational improvements as well as possible applications to other oceanographic quantities of interest.

\section{Background}
\label{sec:background}

\subsection{Argo float profiles}
\label{sec:data}
The Argo program (\cite{Roemmich2009_ArgoProgram}, \cite{Riser2016}, \cite{Wong2020}), part of the Global Ocean Observing System, provides near--global observations of subsurface ocean temperature, salinity, and pressure (and, in some cases, other ocean properties) in the upper 2000 dbar of the ocean using autonomous profiling floats. Here, we use Argo temperature, salinity, and pressure profiles from 2004--2022 obtained from the January 2023 Argo GDAC snapshot \citep{ArgoDataJan2023}. Although the first Argo floats were deployed in 1999, near-global coverage was not achieved until approximately 2004, motivating the use of this start year in many Argo-based mapping products (e.g., \cite{RG2009}, \cite{Hakuba2024}).

A key consideration when mapping OHC is the heterogeneous vertical sampling of Argo profiles. Not all floats provide complete measurements from the surface to 2000 dbar due to instrument limitations, bathymetric constraints, and other technical factors. For example, early Argo floats had reduced sampling capabilities in some regions, particularly in the tropics, limiting observations below 1000 dbar \citep{Wong2020}. In practice, even in the absence of bathymetry and when the float is operating as planned, a many profiles start around 1850 dbar and end 10--15 dbar below the surface. This uneven sampling is particularly important for uncertainty estimation, as observations become increasingly sparse with depth, and motivates jointly mapping multiple pressure layers while accounting for their dependence.

Due to these vertical sampling considerations, we will jointly model OHC in two pressure layers: 15--975 dbar (``upper ocean'') and 975--1850 dbar (``midocean''). The cutoff choice of 975 dbar is motivated by retaining the maximum number of continuous vertical profiles that pass quality control for analysis. After applying the quality control procedure described in \cite{Kuusela2018}, we find 1,540,593 available profiles for the upper ocean layer and 1,128,932 for the midocean layer.

\subsection{Ocean heat content}
\label{sec:problem-def}
For the estimates presented in this paper, we define the ocean heat content (OHC) from 15--1850 dbar at time $t$ within spatial domain $D$ as the time series defined by the integral 
\begin{equation}
\OHC_{15}^{1850}(t) = \underset{{(x,y) \in D}}{\iint} \int_{z=15}^{z=1850} \rho_0 \, c_{p,0} T(x,y,z,t)\,\mathrm{d}z \,\mathrm{d}S(x,y). \label{eq:OHC_def}
\end{equation}
Here $T(x,y,z,t)$ is the potential temperature (i.e., the temperature of a water parcel if brought without heat exchange to the surface) at longitude $x$, latitude $y$, pressure $z$ and time $t$; $\rho_0=1030 \text{ kg/m}^3$ is the average in-situ density of seawater at the surface and  $3989.244 \ \mathrm{J\cdot kg^{-1} \cdot K^{-1}}$ is the average seawater specific heat capacity at the surface (e.g., \cite{potentialenthalpy}). 

Since Argo floats sample at a sufficiently fine resolution over pressure, converting the temperature profiles to potential temperature over $z$ is straightforward using the TEOS-10 toolbox \citep{TEOS-10}. The vertical integral is also easily computed via PCHIP interpolation as described in \cite{univariate}. However, there are two additional considerations for the spatial integral over $D$: first, $T$ is only known at the spatio-temporal locations where a float collected a profile, and second, due to the heterogeneity in vertical sampling, $T$ may not be fully observed for the entirety of the range of $z$. To address these challenges, we will then need to interpolate the vertically integrated temperature profiles for each of the pressure layers defined in Section \ref{sec:data} onto a regular grid to compute the spatial integral. More formally, our goal in this work is to produce accurate monthly estimates and uncertainties for 

\begin{equation}
\begin{split}
\OHC_{1850}^{15}(t) &= \underset{{(x,y) \in D}}{\iint} [\int_{z=15}^{z=975} \rho_0 \, c_{p,0} T(x,y,z,t)\,\mathrm{d}z\,+\int_{z=975}^{z=1850} \rho_0 \, c_{p,0} T(x,y,z,t)\,\mathrm{d}z\,]\mathrm{d}S(x,y)\\&
\approx \sum_{i} [\widetilde{\OHC}_{15}^{975}(x_i,y_i,t)+\widetilde{\OHC}_{975}^{1850}(x_i,y_i,t)]\cdot S_i,
\end{split}\label{eq:OHC_def_xy}
\end{equation}
where $\widetilde{\OHC}_{z_u}^{z_d}(x,y,t)$ is the vertical integral of $T$ from $z_u$ to $z_d$, $(x_i,y_i)$ are points on a $1^\circ \times 1^\circ$ grid over $D$ and $S_i$ is the area of a $1^\circ \times 1^\circ$ grid element whose center point is $(x_i,y_i)$. We define the spatial domain $D = (D_\text{RG}^\up \cap D_\text{data}^\up) \cup (D_\text{RG}^\mi \cap D_\text{data}^\mi)=D^\up \cup D^\mi$ to be the Argo-sampled part of the global ocean after accounting for bathymetry and data availability for each layer. $D_\text{RG}$ is the \cite{RG2009} ocean mask, which excludes polar regions, continental shelves and other shallow areas. $D_\text{data}$ is the domain defined in \cite{univariate} that contains the ocean regions sufficiently well-sampled to estimate the seasonal component of the climatological mean field in a local neighborhood. In summary, $D$ includes the regions where it is feasible to estimate the model parameters for at least one pressure layer (note that when $(x_i,y_i) \in D^\mi \setminus D^\up$, $\widetilde{\OHC}_{15}^{975}(x_i,y_i,t)=0$ and when $(x_i,y_i) \in D^\up \setminus D^\mi$, $\widetilde{\OHC}_{975}^{1850}(x_i,y_i,t)=0$).

The main challenge is obtaining a reliable uncertainty on the estimate of $\OHC^{15}_{1850}(t)$ given the heterogeneity in Argo's vertical sampling. Obtaining an uncertainty on a multivariate functional, such as the global OHC in Equation \eqref{eq:OHC_def_xy}, is not as straightforward as producing a pointwise uncertainty for the OHC estimate at each grid point. For example, \cite{univariate} have previously shown that it is computationally feasible to produce a rigorously quantified uncertainty on the global OHC estimate for a single pressure layer by conditionally simulating from a locally stationary univariate Gaussian process. However, this approach does not provide a full picture of the uncertainty of the sum over pressure layers because it does not account for the vertical spatio-temporal dependence between the layers. In the next subsection, we provide an overview of how the vertical dimension is handled in current Argo data products before proposing our improved mapping and uncertainty quantification framework in Section \ref{sec:methods}.

\subsection{Related work}
\label{sec:related-work}

One of the first data products (i.e., mapped fields) produced after the Argo array achieved sparse global coverage was the Roemmich--Gilson (RG) climatology \citep{RG2009}. The vertical dimension is divided into 58 pressure layers spaced from 10 to 100 dbar apart, onto which the Argo profiles are linearly interpolated. For each pressure layer and month, a weighted least-squares mean field is estimated using the nearest observations from the current and adjacent pressure layer. Then to produce the monthly temperature and salinity anomaly fields, the spatial interpolation is done separately for each month and pressure layer. However, the covariance model only varies by spatial location and does not vary based on month or pressure. One consequential limitation is that it becomes not immediately obvious how to obtain reliable predictive uncertainties on the mapped anomaly fields; this limitation is discussed in further detail by \cite{Kuusela2018}. 

Recent work in the multivariate spatial statistics literature has provided a number of ways to incorporate pressure information into the modeling process. For example, \cite{Yarger2022} and \cite{Korte-Stapff2025} have developed a functional kriging method to predict temperature and salinity from Argo profiles as a function of pressure. This approach is highly flexible since for a given location, the estimated mean function can be evaluated for any pressure from 0-2000 dbar. Modeling the covariance follows a three-step process where functional principal components (FPCs) are first estimated to reduce the high dimensionality in the covariance structure between space, time, and pressure. Next, the FPC bases are used to compute the temperature and salinity scores, which are then jointly modeled using the locally stationary space-time covariance function from \cite{Kuusela2018}. A notable advantage of this functional kriging approach is that pointwise predictive uncertainties for the mapped OHC anomaly field can be obtained for any pressure from 0-2000 dbar. However, obtaining an uncertainty on the global OHC anomaly would require the predictive covariance between every spatial location in the Argo-sampled global ocean, which is not readily available from the local models.

\cite{Salvana2022} handle the vertical dimension by introducing a 3D bivariate model for temperature and salinity with a nonstationary covariance function that varies across latitude, longitude, and pressure. Although the vertical dependence structure can be captured in detail by directly including pressure in the covariance function, this approach also warrants additional computational considerations given the large size of the Argo data. An example of a $400,000 \times 400,000$ local cross-covariance matrix (based on an average of 100 vertical observations per Argo profile) was given, which would be infeasible to store and invert to produce the kriging prediction or estimate the model parameters via maximum likelihood without high-performance computing tools. 

\cite{Bolin2019} jointly mapped Argo temperature fields at 300 dbar and 1500 dbar using a class of stochastic partial differential equation (SPDE) models with normal variance mixture (type G) noise. Along with enabling multivariate modeling of pressure layers, the type G SPDE models are promising in that they can also handle non-Gaussianity, which appears to be more prominent in deeper pressure layers. The scope of the application is, however, limited to a region south of New Zealand due to the stationarity assumption and a subset of three years within the Argo sampling period. In addition, the temporal variation is not modeled.

Finally, \cite{Saduakhas2025} recently expanded on the approaches in \cite{Bolin2019} and \cite{Kuusela2018} to jointly map Argo temperature and salinity using local hierarchical multivariate Matérn-SPDE models with correlated nugget effects. This approach provides an improved representation of fine-scale ocean processes, which also leads to benefits for uncertainty quantification. Although this is a significant step in rigorous uncertainty quantification for multivariate Argo fields, the temporal dependence structure is not considered which is necessary, for example, to produce uncertainties on OHC trend estimates.

Our goal in this paper is therefore to contribute a mapping and uncertainty quantification framework that incorporates the vertical spatio-temporal dependence structure while providing statistically rigorous, computationally feasible uncertainty estimates for global OHC. As a starting point, we will first describe a recent univariate framework from \cite{univariate} where the vertically integrated temperature residuals in each pressure layer are mapped separately.

\section{Bivariate mapping and uncertainty quantification}
\label{sec:methods}

\subsection{Motivation for bivariate extension}
\label{sec:univariate-model}
\cite{univariate} recently developed a mapping and uncertainty quantification framework based on the statistical model
\begin{equation}
\widetilde{\OHC}_l(x,y,t) = \mu (x,y,t) + a(x,y,t) + \varepsilon(x,y,t), \label{eq:model_def}
\end{equation}
for a single pressure layer $l$, where the climatological mean field $\mu(x,y,t)$ is estimated by fitting local polynomial regression functions on a regular grid over $D^l$. The anomaly field $a(x,y,t)$ is then modeled for each point on a spatio-temporal grid $D^l \times D^t$ using a zero-mean locally stationary Gaussian process, where the additive Gaussian nugget effect $\varepsilon(x,y,t)$ is assumed independent over space and time within the local window. 

Let us now consider the uncertainty of the total $\OHC_{15}^{1850}(t)=\OHC_{15}^{975}(t)+\OHC_{975}^{1850}(t)$ in Eq. \eqref{eq:OHC_def_xy}. More formally, we need to produce an estimate of the conditional variance
\begin{equation}
\begin{split}
\Var(\OHC_{15}^{1850}(t)|\text{data})=&\Var(\OHC_{15}^{975}(t)|\text{data})+\Var(\OHC_{975}^{1850}(t)|\text{data})\\
&+2\text{Cov}(\OHC_{15}^{975}(t),\OHC_{975}^{1850}(t)|\text{data}).
\end{split}
\label{eq:cond-var-total}
\end{equation}
\cite{univariate} developed a local conditional simulation algorithm based on the fact that under certain conditions, Gaussian processes can be expressed as convolutions of Gaussian white noise (e.g., \cite{Higdon2002}). Using this idea, we can separately simulate from the conditional distributions of $\{\widetilde{\OHC}_{15}^{975}(x_i,y_i,t_i)|\}_i$ and $\{\widetilde{\OHC}_{975}^{1850}(x_i,y_i,t_i)\}_i$ by convolving white noise within a local window centered on grid point $(x_i,y_i,t_i)$ with a kernel computed from the local predictive covariance matrix. The estimates of $\Var(\OHC_{975}^{15}(t)|\text{data})$ and $\Var(\OHC_{1850}^{975}(t)|\text{data})$ then follow by substituting the respective conditional simulation realizations into the single pressure layer analogue of the summation in Equation \eqref{eq:OHC_def_xy} and computing the corresponding sample variances. However, note that this approach does not include the effect of the conditional covariance term on the right-hand side of Equation \eqref{eq:cond-var-total}. As a result, \cite{univariate} produces a conservative upper bound on the total OHC uncertainty, using
\begin{equation}
\Var(\OHC_{15}^{1850}(t)|\text{data})\leq \left(\sqrt{\Var(\OHC_{15}^{975}(t)|\text{data})}+\sqrt{\Var(\OHC_{975}^{1850}(t)|\text{data})}\right)^2.
\label{eq:conservative-bound}
\end{equation}

This indicates that we can improve the total OHC uncertainty estimates by modeling the vertical spatio-temporal dependence between the two pressure layers. In addition, since the midocean layer is more sparsely sampled, we expect the OHC estimates to improve by borrowing strength from the more densely sampled upper ocean layer. In the next section, we introduce bivariate extensions of the locally stationary Gaussian process model and local conditional simulation algorithm that incorporate this dependence.

\subsection{Mapping: bivariate locally stationary GP regression
}
\label{sec:local-gp}
In the bivariate extension of \cite{univariate}, we jointly model the upper ocean and midocean OHC as specified below:
\begin{equation}
\begin{bmatrix}
\widetilde{\OHC}_{15}^{975}(x,y,t)\\
\widetilde{\OHC}_{975}^{1850}(x,y,t)
\end{bmatrix} = 
\begin{bmatrix}\mu_{\text{up}}(x,y,t)\\
\mu_{\text{mid}}(x,y,t)
\end{bmatrix}+ 
\begin{bmatrix}a_{\text{up}}(x,y,t)\\
a_{\text{mid}}(x,y,t)
\end{bmatrix}+
\begin{bmatrix}\varepsilon_{\text{up}}(x,y,t)\\
\varepsilon_{\text{mid}}(x,y,t)
\end{bmatrix}=\boldsymbol{\mu}+\boldsymbol{a}+\boldsymbol{\varepsilon}. \label{eq:model_def_bivariate}
\end{equation}
In this section, we focus on modeling the zero-mean residuals
\begin{equation} 
    \boldsymbol{r}(x,y,t):=
    \begin{bmatrix}
    \widetilde{\OHC}_{15}^{975}(x,y,t)\\
    \widetilde{\OHC}_{975}^{1850}(x,y,t)
    \end{bmatrix} - 
    \begin{bmatrix}\mu_{\text{up}}(x,y,t)\\
    \mu_{\text{mid}}(x,y,t)
    \end{bmatrix}= 
    \begin{bmatrix}a_{\text{up}}(x,y,t)\\
    a_{\text{mid}}(x,y,t)
    \end{bmatrix}+
    \begin{bmatrix}\varepsilon_{\text{up}}(x,y,t)\\
    \varepsilon_{\text{mid}}(x,y,t)
    \end{bmatrix}.   
    \label{eq:res_def_bivariate}
\end{equation}
The estimated bivariate mean field $\boldsymbol{\hat\mu}=[\hat \mu_\up, \hat \mu_\mi]^\T$ at an observed profile location $(x_p,y_p,t_p)$ is straightforward to compute as the estimated mean field in \cite{univariate} at the closest spatial grid point $(x^*,y^*)$ applied to each pressure layer separately. This leads us to define the observed residual vector as $\boldsymbol{\hat r}(x_p,y_p,t_p)=[\widetilde{\OHC}_{15}^{975}(x_p,y_p,t_p), 
\widetilde{\OHC}_{975}^{1850}(x_p,y_p,t_p)]-[\hat \mu_\up(x^*,y^*,t_p), \hat \mu_\mi(x^*,y^*,t_p)]$. However, to model the stochastic terms $\boldsymbol{a}$ and $\boldsymbol{\varepsilon}$, we will need to take the vertical spatio-temporal dependence between the two layers into consideration.

As in \cite{univariate}, let us define a window
\begin{equation}
\widetilde{W}(x^*,y^*,t^*)=[x^*-\lambda_G,x^*+\lambda_G] \times [y^*-\lambda_G,y^*+\lambda_G] \times [t^*-\lambda_t,t^*+\lambda_t]
\end{equation}
centered on a gridpoint $(x^*,y^*,t^*)$ on a uniform $1^\circ \times 1^\circ \times 1$ month grid over the spatio-temporal domain $D \times D^t$, where $D$ is the spatial domain defined in Section \ref{sec:problem-def} and $D^t$ is the time domain of interest within the Argo sampling period. $\lambda_G$ and $\lambda_t$ are hyperparameters defining the window size over space in degrees and over time in months respectively. Let us now consider the set of all such windows over $D^t$
\begin{equation}
W_{\lambda_{G},\lambda_{t}} (x^*,y^*,\tau^*)= \{ \widetilde{\W}_{\lambda_{G},\lambda_{t}} (x^*,y^*,t^*) \::\: t^* \in D_t, \: t^* \bmod 365 = \tau^* \}, \end{equation}
where $\tau^* \in [0,365]$ is a yearday on the same temporal grid as $t^*$. 

Then, for a given $W$, let us now assume that the OHC anomaly process for year $i$ is a mean-zero bivariate Gaussian process 
\begin{equation}
\label{eq:gp_bivariate}
\begin{bmatrix}
a_{\text{up},i}\\
a_{\text{mid},i}
\end{bmatrix}
 \iid \text{GP} (
 [0,0]^\T,
 \boldsymbol{K}(\boldsymbol{z}_1,\boldsymbol{z}_2;\boldsymbol{\theta}_W)),
\end{equation}
where $\boldsymbol{K}$ is a $2 \times 2$ matrix-valued stationary covariance function, $\boldsymbol{z}=(x,y,t)$ is a spatio-temporal location,
\begin{equation}
\boldsymbol{\theta}_W=(\phi_{W,\up},\phi_{W,\mi},\theta_{W,\lat,\up},\theta_{W,\lat,\mi},\theta_{W,\lon,\up},\theta_{W,\lon,\mi},\theta_{W,t,\up},\theta_{W,t,\mi},\beta_W)^\T
\end{equation}
is a vector of covariance parameters for $W$ (where the individual parameters are defined below) and the i.i.d.~assumption is across the years in $W$. For the diagonal elements of K representing the spatio-temporal dependence within a single pressure layer $l$, we use the stationary anisotropic exponential covariance function $K_{ll}(\bm{z}_1,\bm{z}_2;\bm{\theta}_W)=\phi_{W,l}\exp(-d(\boldsymbol{z}_1,\boldsymbol{z}_2))$ where $d$ is the distance function defined in \cite{Kuusela2018}
\begin{equation}
    d(\bm{z}_1,\bm{z}_2) = \sqrt{(\bm{z}_1-\bm{z_2})^\T\bm{\Theta}_l^{-1}(\bm{z}_1-\bm{z}_2)}.
    \label{eq:dist_def}
\end{equation}
Here, $\bm{\Theta}_{lm}=\text{diag}(\theta_{W,\text{lat,l}}^2,\theta_{W,\text{lon,l}}^2,\theta_{W,t,l}^2)$ where $(\theta_{W,\text{lat,l}},\theta_{W,\text{lon,l}},\theta_{W,t,l})^\T$ are the latitude, longitude, and time length scale parameters for layer $l$.

To model the vertical spatio-temporal dependence between two unique pressure layers $l$ and $m$, we use the anisotropic cross-covariance function introduced in \cite{Kleiber2012}, which in the case of exponential marginal fields, is given by
\begin{equation}
{K_{lm}}(\bm{z}_1,\bm{z}_2;\bm{\theta}_W)=  \beta_W\frac{\delta_l\delta_m}{|\bm{\Theta}_{lm}|^{1/2}}\exp{\left(-\sqrt{(\bm{z_1-z_2})^{\text{T}}\bm{\Theta}_{lm}^{-1}(\bm{z_1-z_2}}) \right)}
\end{equation} 
where $\delta^2_i=\phi_{W,i}\sqrt{|\text{diag}(\theta_{W,\text{lat}}^2,\theta_{W,\text{lon}}^2,\theta_{W,t}^2)|}$, $\beta_W$ is the vertical cross-correlation within window $W$, and $\bm{\Theta}_{lm}=\frac{1}{2}\left(\text{diag}(\theta_{W,\text{lat},l}^2,\theta_{W,\text{lon},l}^2,\theta_{W,t,l}^2)+\text{diag}(\theta_{W,\text{lat,m}}^2,\theta_{W,\text{lon,m}}^2,\theta_{W,t,m}^2)\right)$. 

Additionally, following the model definition in \eqref{eq:model_def_bivariate}, we define an additive bivariate Gaussian nugget effect
\begin{equation}
\begin{bmatrix}
\varepsilon_{\text{up}}\\
\varepsilon_{\text{mid}}
\end{bmatrix}
 \iid \text{MVN} \left(
 \begin{bmatrix}
 0\\
 0
 \end{bmatrix},
 \begin{bmatrix}
 \sigma_{W,\up}^2 & \rho_W\sigma_{W,\up}\sigma_{W,\mi}\\
 \rho_W\sigma_{W,\up}\sigma_{W,\up} & \sigma_{W,\mi}^2
 \end{bmatrix}\right)
\end{equation}
where $\sigma_W^2$ is the marginal nugget variance for layer $i$, $\rho_W \in [-1,1]$ is the nugget vertical correlation, and the i.i.d.~assumption is over space and time within W. 

We use the residuals $\bm{\hat r}$ in $W$ to numerically find the maximum likelihood estimates (MLEs) of the model parameters for each $W$. In practice, for each pressure layer separately, we first find the MLEs of the univariate model parameters for each of the 12 months of the year,  as in \cite{univariate}, and use the gridpoint-wise annual medians as the plug-in estimates for the marginal covariance and nugget parameters. We find that this choice makes finding the MLEs of the cross-correlation parameters $(\beta_W, \rho_W)$ a simpler optimization problem that is more numerically stable. This also simplifies comparison with the univariate approach since the marginal models are exactly the same. We also use the annual medians of the monthly MLEs of $(\beta_W, \rho_W)$ as the final plug-in estimates. Since any finite subset of a Gaussian process follows a multivariate Gaussian distribution, the point prediction (i.e., conditional mean) of the residual vector $\boldsymbol{r}(x^*,y^*,t^*)$ at a grid point $(x^*,y^*,t^*)$ and the associated predictive uncertainty (i.e., conditional variance) then follow as the bivariate kriging mean and variance (e.g., Chapter 28 of \cite{handbook}) computed based on the plug-in estimates of the model parameters and the mean-centered observed residuals $\boldsymbol{\hat r}$ within the window $\widetilde{\W}_{\lambda_{G},\lambda_{t}} (x^*,y^*,t^*)$. Then, following the model definition in \eqref{eq:res_def_bivariate}, it simply remains to add back the mean field estimate vector $\boldsymbol{\hat \mu}(x^*,y^*,t^*)$ to obtain the final point prediction $[\widehat{\widetilde{\OHC}}_{15}^{975}(x^*,y^*,t^*), 
\widehat{\widetilde{\OHC}}_{975}^{1850}(x^*,y^*,t^*)]^\T$ of $[\widetilde{\OHC}_{15}^{975}(x^*,y^*,t^*), 
\widetilde{\OHC}_{975}^{1850}(x^*,y^*,t^*)]^\T$ for each gridpoint $(x^*,y^*,t^*)$.

\subsection{Uncertainties: bivariate local conditional simulation}
\label{sec:condsim}
As mentioned in the previous subsection, the pointwise predictive uncertainty (conditional variance) of $[\widetilde{\OHC}_{15}^{975}(x_i,y_i,t_i), \widetilde{\OHC}_{975}^{1850}(x_i,y_i,t_i)]^\T$ at a gridpoint $i$ is easily obtained as the kriging variance. However, to obtain an uncertainty of the total $\OHC_{15}^{1850}(t)$ in Equation \eqref{eq:OHC_def_xy}, we would require the conditional covariance between all gridpoints in $D^l \times D^t$  for each pressure layer $l$, which would be infeasible to store and compute. Moreover, we would also require the conditional covariance between $\OHC_{15}^{1850}(t)$ and $\OHC_{975}^{1850}(t)$, which does not follow immediately from the above model.

We will instead simulate from the predictive distribution of the bivariate model, which will incorporate the effect of this conditional covariance between pressure layers. \cite{univariate} developed a local conditional simulation algorithm to simulate from the marginal conditional distribution $p(\{\widetilde{\OHC}_l(x_{i},y_{i},t_{i})\}_i|\text{data}\})$ for a single pressure layer $l$. The main idea of this algorithm comes from the fact that since $\widetilde{\OHC}_l(x_{i},y_{i},t_{i})\}_i|\text{data}$ is a Gaussian process, we can rewrite it as the convolution
\begin{equation}
    f(\s,t) = \int_{\widetilde{\W}_{\lambda_{G},\lambda_{t}} (\s,t)} h_{\s,t}(\u,v) w(\u,v) \,\mathrm{d}\u \,\mathrm{d} v,  \quad (\s,t) \in D \times D_t,
\end{equation}
where $\widetilde{\W}_{\lambda_{G},\lambda_{t}}$ is the local window in Section \ref{sec:local-gp} centered on spatio-temporal location ($\bm{s},t$), $h_{\s,t}(\u,v)$ is a kernel and $w(\u,v)$ is a Gaussian white noise process. This expression motivates an algorithm on a spatio-temporal grid over $D \times D_t$ where (1) we simulate $w$ on the grid and keep fixed, (2) we find the local conditional covariance matrix and its symmetric square root (i.e., kernel) for the window centered on grid point $i$, and (3) we convolve $w$ with the kernel and store the center point to obtain the conditional simulation realization for grid point $i$. We explain the bivariate extension of this algorithm in greater detail below.  

We first simulate Gaussian white noise on a $1^\circ \times 1^\circ \times 1$ month grid for each of the two pressure layers separately. We then use the fact that the conditional distribution of a multivariate Gaussian random variable is also multivariate Gaussian to compute the local predictive bivariate covariance matrix for the window $\widetilde{\mathcal{W}}_{\lambda_{G},\lambda_t,i}$ centered on grid point $i$, which is given by
\begin{equation}
\boldsymbol{\Sigma}_i=\boldsymbol{\Sigma}_{Z^*}(\boldsymbol{\hat \theta}_i)-\boldsymbol{\Sigma}_{Z^*Z_p}(\boldsymbol{\hat \theta}_i)[\boldsymbol{\Sigma}_{Z_p}(\boldsymbol{\hat \theta}_i)+\boldsymbol{\Sigma}_{\varepsilon}(\hat \sigma_{i,\up},\hat \sigma_{i,\mi},\hat\rho_i)]^{-1}\boldsymbol{\Sigma}_{Z^*Z_p}(\boldsymbol{\hat \theta}_i)^{\text{T}},
\end{equation}
where $Z^*$ is the set of all grid points $\{\boldsymbol{z}_\up^*=(x^*_\up,y^*_\up,t^*_\up)$, $\boldsymbol{z}_\mi^*=(x^*_\mi,y^*_\mi,t^*_\mi)\}$and $Z_p$ is the set of all observed profile locations $\{\boldsymbol{z}_{\up,p}=(x_{\up,p},y_{\up,p},t_{\up,p})$, $\boldsymbol{z}_{\mi,p}=(x_{\mi,p},y_{\mi,p},\\t_{\mi,p})\}$, respectively, in  $\widetilde{\mathcal{W}}_{\lambda_{G},\lambda_t,i}$; and $(\boldsymbol{\hat \theta}_i, \hat \sigma_{i,\up},\hat \sigma_{i,\mi},\hat\rho_i)$ are the plug-in model parameter estimates from Section \ref{sec:local-gp} for grid point $i$.
Then, with a slight abuse of notation, $\boldsymbol{\Sigma}_{Z^*}=\boldsymbol{K}(Z^*,Z^*;\boldsymbol{\hat \theta}_i),
\boldsymbol{\Sigma}_{Z_p}=\boldsymbol{K}(Z_p,Z_p;\boldsymbol{\hat \theta}_i),
$ and $\boldsymbol{\Sigma}_{Z^*Z_p}=\boldsymbol{K}(Z^*,Z_p;\boldsymbol{\hat \theta}_i)$ are the covariance matrices obtained by evaluating the covariance function $\boldsymbol{K}$ in \eqref{eq:gp_bivariate} on all pairs of spatio-temporal locations $(\boldsymbol{z}^*_1,\boldsymbol{z}^*_2)$ in $Z^*$, $(\boldsymbol{z}_{p,1},\boldsymbol{z}_{p,2})$ in $Z_p$, and $(\boldsymbol{z}^*,\boldsymbol{z}_p)$ with $\boldsymbol{z}^* \in Z^*$ and $\boldsymbol{z}_p \in Z_p$ respectively. Note that since the nugget is independent over space and time within $\widetilde{\mathcal{W}}_{\lambda_{G},\lambda_t,i}$, the marginal nugget covariance matrix $\boldsymbol{\Sigma
}_{\varepsilon,ll}$ is simply $\hat\sigma^2_{i,l}\boldsymbol{I}$ and the nugget cross-covariance matrix $\boldsymbol{\Sigma
}_{\varepsilon,lm}$ has entries $\hat\rho_i \hat\sigma_{i,l}
\hat\sigma_{i,m}$ for the locations where both pressure layers are observed and is otherwise zero. Following step (2) in the algorithm above, we then compute the eigendecomposition of $\boldsymbol{\Sigma}_i$ to obtain the symmetric square root $\boldsymbol{\Sigma}_i^{1/2}$. However, the bivariate case requires extra care to ensure the simulated nuggets include the vertical dependence. That is, the bivariate conditional simulation realization for grid point $i$ is given by
\begin{equation}
\begin{bmatrix}
    \widetilde{\OHC}_{15,\text{sim}}^{975}(x_{i},y_{i},t_i)\\
    \widetilde{\OHC}_{975,\text{sim}}^{1850}(x_{i},y_{i},t_i)
\end{bmatrix} = 
\begin{bmatrix}
    \widehat{\widetilde{\OHC}}_{15}^{975}(x_i,y_i,t_i)\\
    \widehat{\widetilde{\OHC}}_{975}^{1850}(x_i,y_i,t_i)
\end{bmatrix}+
\begin{bmatrix}
    f_{i,\up}\\
    f_{i,\mi}
\end{bmatrix}+
\begin{bmatrix}
    \hat\sigma_{i,\up}w_{i,\up}\\
    aw_{i,\up}+b_{i,\mi}w_{i,\mi}
\end{bmatrix},
\end{equation}
where the $\boldsymbol{\widehat{\widetilde{\OHC}}}$ vector contains the point predictions from Section \ref{sec:local-gp} and the $\boldsymbol{f}$ vector contains the center point values of $\boldsymbol{\Sigma}_i^{1/2}$ multiplied by the spatial white noise which captures both the vertical and spatio-temporal dependence in $f_{i,up}$ and $f_{i,\mi}$; $a=\hat\rho_i\hat\sigma_{i,\mi}$ and $b=\hat\sigma_{i,\mi}\sqrt{1-\hat\rho_i^2}$ are constants computed from the estimated nugget parameters; and $w_{i,\up}$ and $w_{i,\mi}$ are independent Gaussian white noise realizations for grid point $i$ (note that these are a different set of white noise realizations from those used to obtain $\boldsymbol{f}$).

Since we jointly conditionally simulate the two fields, the effect of the conditional covariance term in Equation \eqref{eq:cond-var-total} propagates into the simulations. This means that we can form an improved uncertainty estimate for the total OHC where we simply substitute $\{\widetilde{\OHC}_{15,\text{sim}}^{975}(x_{i},y_{i},t), \widetilde{\OHC}_{975,\text{sim}}^{1850}(x_{i},y_{i},t)\}_i$ into Equation \eqref{eq:OHC_def_xy} for every realization and compute the sample variance. This carries the benefit from the univariate approach where the dependence over time $t$ is also captured. We show in the following sections that this results in a substantial reduction in uncertainty and enables quantifying uncertainties for downstream oceanographic quantities that were not possible using the previous univariate approach.

\section{Validation: effect of modeling the vertical dimension}
\label{sec:validation}

In Sections \ref{sec:validation} and \ref{sec:results}, we compare validation results and OHC uncertainties obtained using the approach in this paper (``bivariate model'') with the approach in \cite{univariate} (``univariate model''). In all comparisons below, we apply both mapping frameworks to Argo temperature profiles from 2004--2022 and set the mean field window size to $\lambda=5$ degrees, the Gaussian process spatial window size to $\lambda_G=10$ degrees, time window size to $\lambda_{t}=1.5$ months (note that these parameters correspond to half the window width). We set the number of conditional simulation realizations to 500 and the random seed used to generate white noise to be the same as in \cite{univariate}. We estimate the model parameters using BFGS optimization implemented in the MATLAB Optimization Toolbox \citep{matlab-optim}. All computational tasks, including parameter estimation and conditional simulation, were facilitated by the Bridges-2 supercomputer \citep{Brown2021-ji} at the Pittsburgh Supercomputing Center (PSC). 

\subsection{Point predictions of ocean heat content anomalies}
\begin{figure}[!h]
    \centering
    \subfigure[Upper ocean, LOOO]{
        \includegraphics[width=6cm,trim = 0.5cm 0cm 1cm 0cm,clip = true]{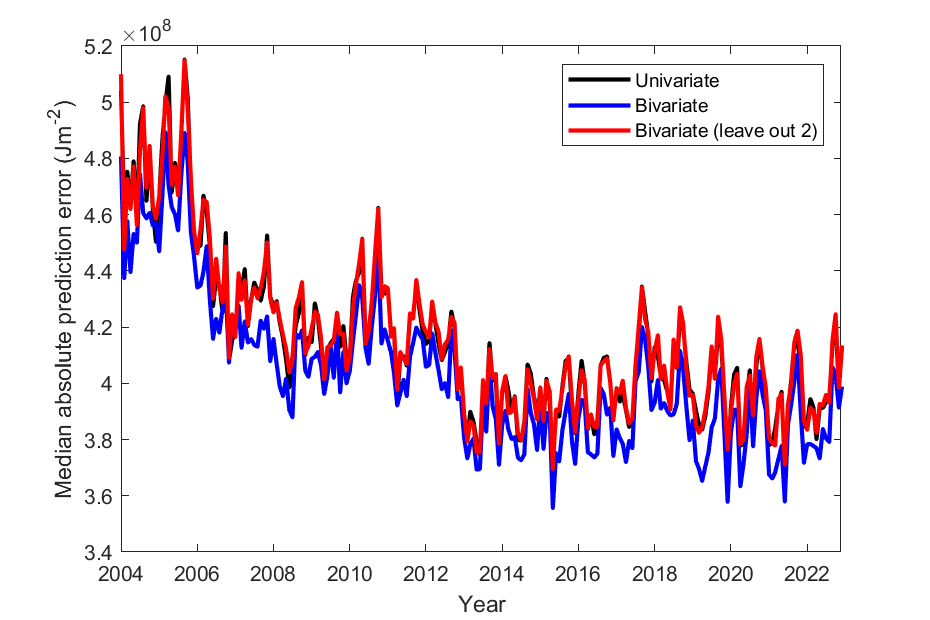}}
    \subfigure[Midocean, LOOO]{
        \includegraphics[width=6cm,trim = 0.5cm 0cm 1cm 0cm,clip = true]{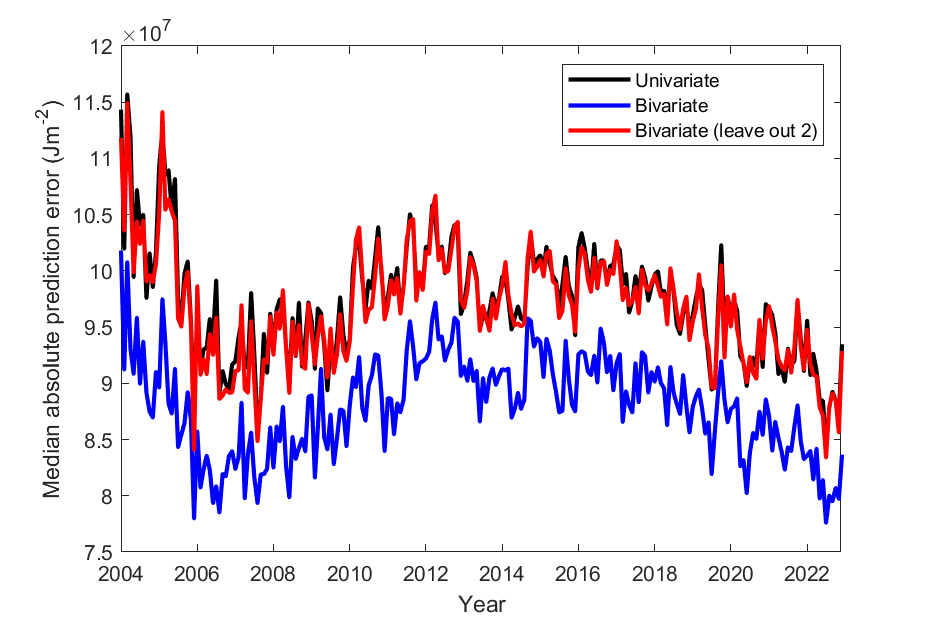}}\\
    \subfigure[Upper ocean, LOFO]{
        \includegraphics[width=6cm,trim = 0.5cm 0cm 1cm 0cm,clip = true]{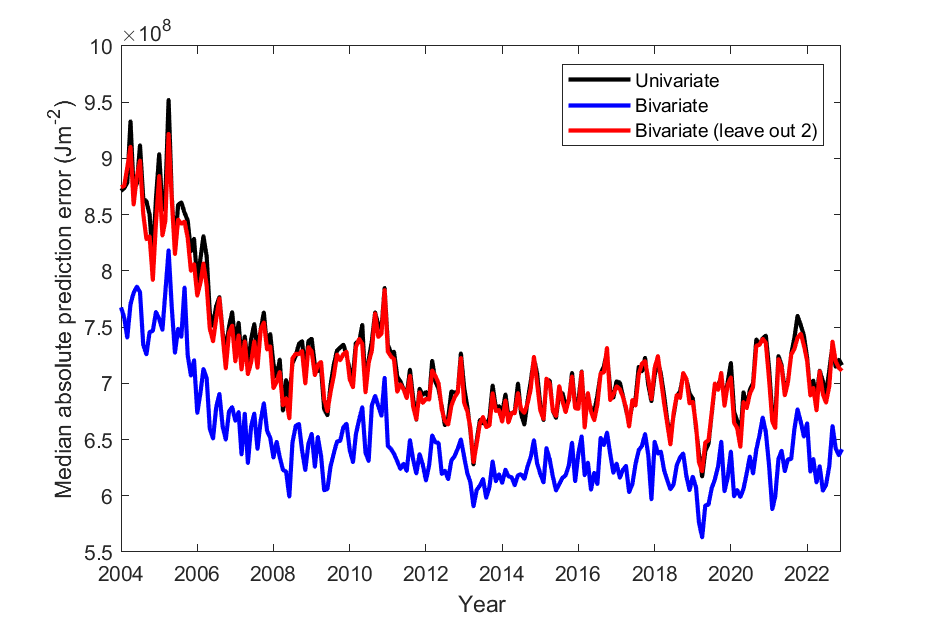}}
    \subfigure[Midocean, LOFO]{
        \includegraphics[width=6cm,trim = 0.5cm 0cm 1cm 0cm,clip = true]{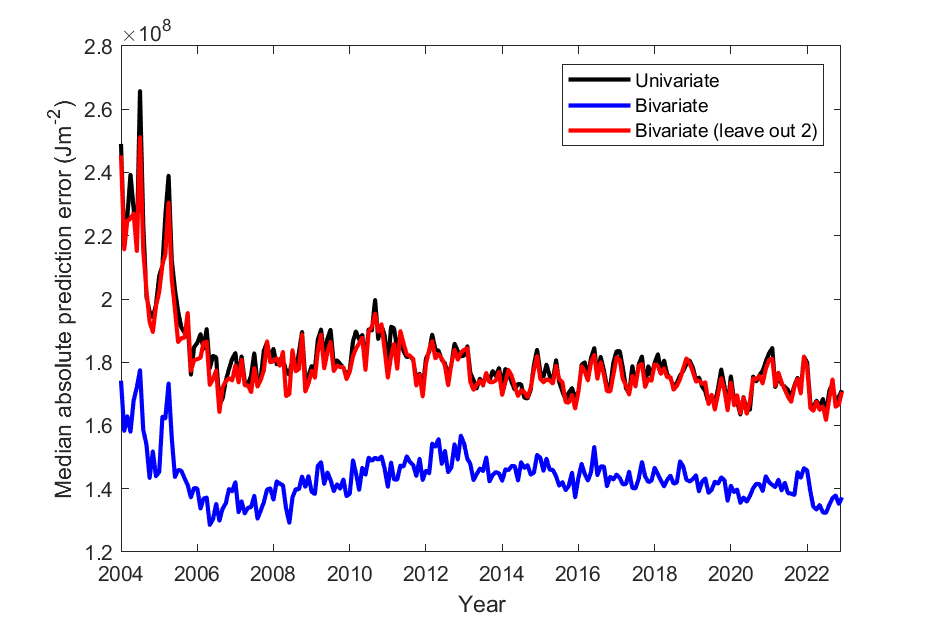}}\\
    \caption{Comparison of cross-validated monthly median absolute prediction errors between univariate and bivariate models.}
    \label{fig:cv_median}
\end{figure}

We first evaluate the effect of modeling the vertical dimension on the point predictions. In Figure \ref{fig:cv_median}, we validate the OHC anomaly point predictions using leave-one-observation-out (LOOO; top row) and leave-one-float-out (LOFO; bottom row) cross-validation. The distinction between these cross-validation schemes is explained in greater detail in \cite{Kuusela2018}. Note that validating the bivariate model motivates two cases where we leave out both pressure layers (red line) or only the pressure layer where a point prediction is desired (blue line). While the former evaluates predictive performance at a fully unobserved location/neighborhood, the latter can reveal whether observing one pressure layer provides useful information for predicting the other. We observe that when we leave out both pressure layers, the bivariate model has comparable prediction errors to the univariate model (black line) for both LOFO and LOOO cross-validation. However, notice that observing the opposite pressure layer from the prediction (blue line) tends to result in reduced prediction errors, most prominently in the midocean (right column). This confirms that in the bivariate model, the midocean is able to borrow predictive strength from the more densely-observed upper ocean. We also see that the upper ocean is able to derive some but less pronounced benefit from observing the midocean.

\label{sec:cv-kriging}

\subsection{Pointwise uncertainties}
\begin{figure}[!h]
    \centering
    \subfigure[Upper ocean]{
        \includegraphics[width=7.5cm,trim = 1.5cm 3.5cm 0.8cm 3cm,clip = true]{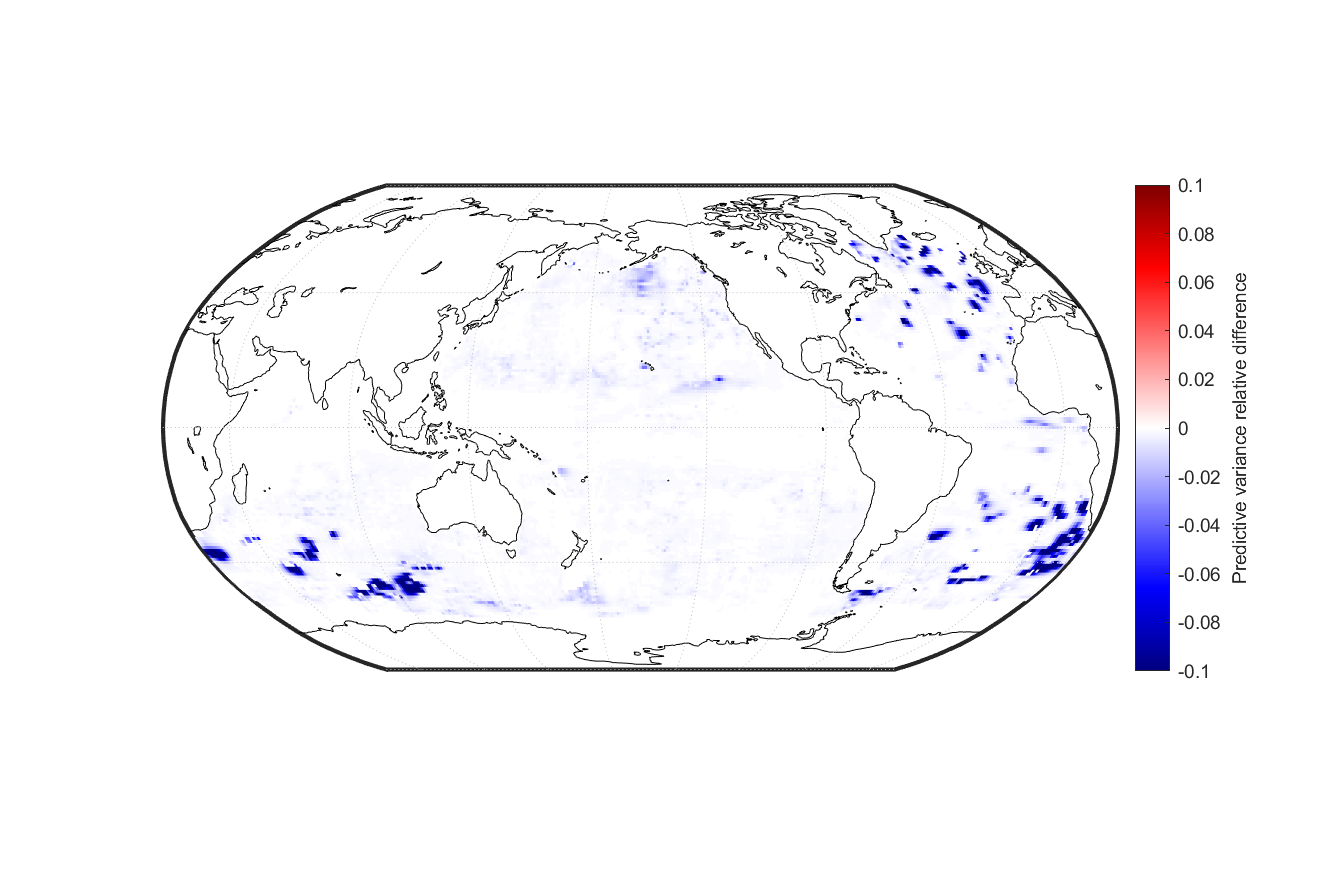}}
    \subfigure[Midocean]{\includegraphics[width=7.5cm,trim = 1.5cm 3.5cm 0.8cm 3cm,clip = true]{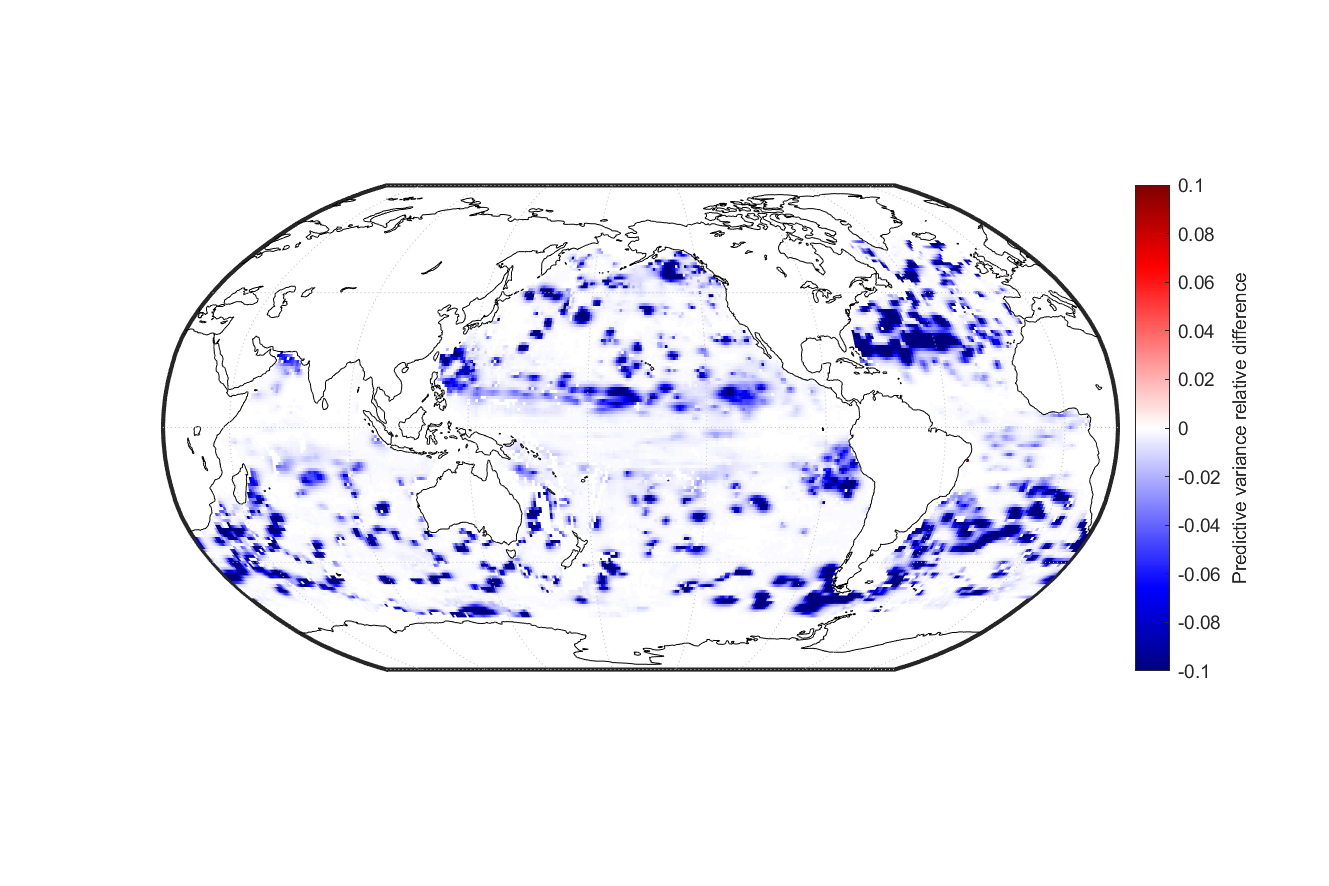}}
    \caption{Relative difference in predictive variance between the univariate and bivariate mapping approaches (February 2010).}
    \label{fig:var_diff}
\end{figure}

Next, we compute the relative difference in predictive variance between the two models, which we define for a spatio-temporal grid point $(x^*,y^*,t^*)$ as 
\begin{equation}
    \frac{\sigma^{2}(x^*,y^*,t^*)-\sigma^{'2}(x^*,y^*,t^*)}{\sigma^{'2}(x^*,y^*,t^*)}
\end{equation} where $\sigma^{2}$ refers to the bivariate model predictive variance in Section \ref{sec:local-gp} and $\sigma^{'2}$ refers to the univariate model predictive variance. A map of this relative difference for February 2010 is shown in Figure \ref{fig:var_diff}. We see that the bivariate model has similar (equal predictive variance; white) or improved (negative relative difference; blue) predictive performance compared to the univariate model. Depending on the ocean region, this improvement can be as much as 10\% or more. The predictive variance reduction is overall larger for the midocean, indicated by the higher intensity and frequency of blue regions. Combined with the above cross-validation results, this suggests that the bivariate model provides a substantial improvement in predictive performance for midocean OHC. Regions with largest improvement (e.g., the Gulf Stream region) tend to correspond to where the total correlation between the pressure layers is stronger and more positive (Figure \ref{fig:total-corr} in Section~\ref{sec:param-estimates}).

\subsection{Conditional spatio-temporal dependence}
\label{sec:cv-condsim}

Here, we evaluate the effect of modeling vertical dependence on the conditional simulations. We will extend the paired leave-one-float-out (LOFO) cross-validation procedure in \cite{univariate} to the bivariate model. Let $\{(x_1,y_1,t_1,f_1),(x_2,y_2,t_2,f_2)\}$ denote a pair of spatio-temporal locations where a vertical profile was collected, where $f_1$ and $f_2$ are the ID numbers of the Argo floats that collected the respective profiles. Now, let $\{(x^*_1,y^*_1,t_1^*),(x^*_2,y^*_2,t_2^*)\}$ be the corresponding closest spatio-temporal grid points. For pressure layers $l$ and $m$, we produce conditional simulation realizations 
\begin{equation}
\{\widetilde{\OHC}_{l,\text{sim}{,-(f_1, f_2)}}(x_1^*,y_1^*,t_1^*),\widetilde{\OHC}_{m,\text{sim}{,-(f_1, f_2)}}(x_2^*,y_2^*,t_2^*)\},
\end{equation}point predictions $\{\widehat{\widetilde{\OHC}}_{l,{-(f_1, f_2)}}(x_1^*,y_1^*,t_1^*),\widehat{\widetilde{\OHC}}_{m,{-(f_1, f_2)}}(x_2^*,y_2^*,t_2^*)\}$, and the corresponding predictive variances $\{\sigma^2_{l,-(f_1,f_2)}(x_1^*,y_1^*,t_1^*),\sigma^2_{m,-(f_1,f_2)}(x_2^*,y_2^*,t_2^*)\}$. We expect the conditionally simulated standardized pair

\begin{tiny}
\begin{equation}
    \left(\frac{\widetilde{\OHC}_{l,\text{sim}{,-(f_1, f_2)}}(x^*_1,y^*_1,t^*_1)-\widehat{\widetilde{\OHC}}_{l,-(f_1, f_2)}(x^*_1,y^*_1,t^*_1)}{\sigma_{l,-(f_1, f_2)}(x^*_1,y^*_1,t^*_1)},\frac{\widetilde{\OHC}_{m,\text{sim}{,-(f_1, f_2)}}(x^*_2,y^*_2,t^*_2)-\widehat{\widetilde{\OHC}}_{m,-(f_1, f_2)}(x^*_2,y^*_2,t^*_2)}{\sigma_{m,-(f_1, f_2)}(x^*_2,y^*_2,t^*_2)}\right)
\end{equation}
\end{tiny}
and the observed standardized pair
\begin{tiny}
\begin{equation}
    \left(\frac{\widetilde{\OHC}_l(x^*_1,y^*_1,t^*_1)-\widehat{\widetilde{\OHC}}_{l,-(f_1,f_2)}(x^*_1,y^*_1,t^*_1)}{\sigma_{l,-(f_1,f_2)}(x^*_1,y^*_1,t^*_1)},\frac{\widetilde{\OHC}_m(x^*_2,y^*_2,t^*_2)-\widehat{\widetilde{\OHC}}_{m,-(f_1,f_2)}(x^*_2,y^*_2,t^*_2)}{\sigma_{m,-(f_1,f_2)}(x^*_2,y^*_2,t^*_2)}\right)
\end{equation}
\end{tiny}to follow standard bivariate Gaussian distributions with the same correlation \\$\rho_{-(f_1,f_2)}(x^*_1,y^*_1,t^*_1,x^*_2,y^*_2,t^*_2)$. We can therefore assess the model's performance in capturing the observed conditional spatio-temporal dependence by comparing the empirical Pearson correlation coefficients of the conditionally simulated and observed standardized pairs in bins of spatio-temporal lag. Note that when $l=m$, this is a straightforward extension of the procedure in \cite{univariate} where we validate the conditional dependence within a single pressure layer. Otherwise, when $l \neq m$, we now additionally validate the vertical dependence between the two pressure layers.

\begin{figure}[!h]
    \centering
    \subfigure[Upper ocean (univariate)]{
        \includegraphics[width=7.5cm,trim = 0.5cm 0cm 1cm 0.5cm,clip = true]{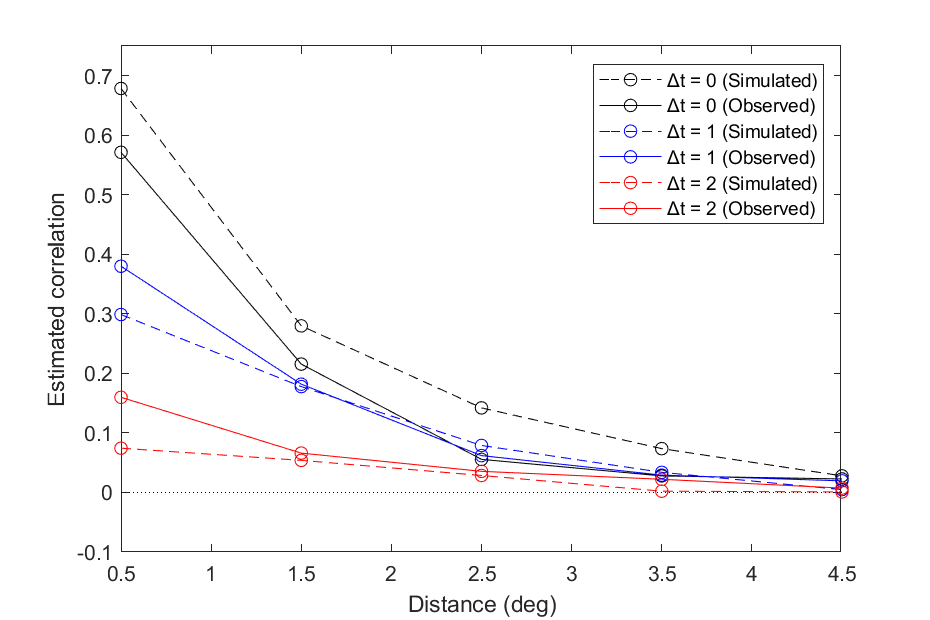}}
    \subfigure[Upper ocean (bivariate)]{
        \includegraphics[width=7.5cm,trim = 0.5cm 0cm 1cm 0.5cm,clip = true]{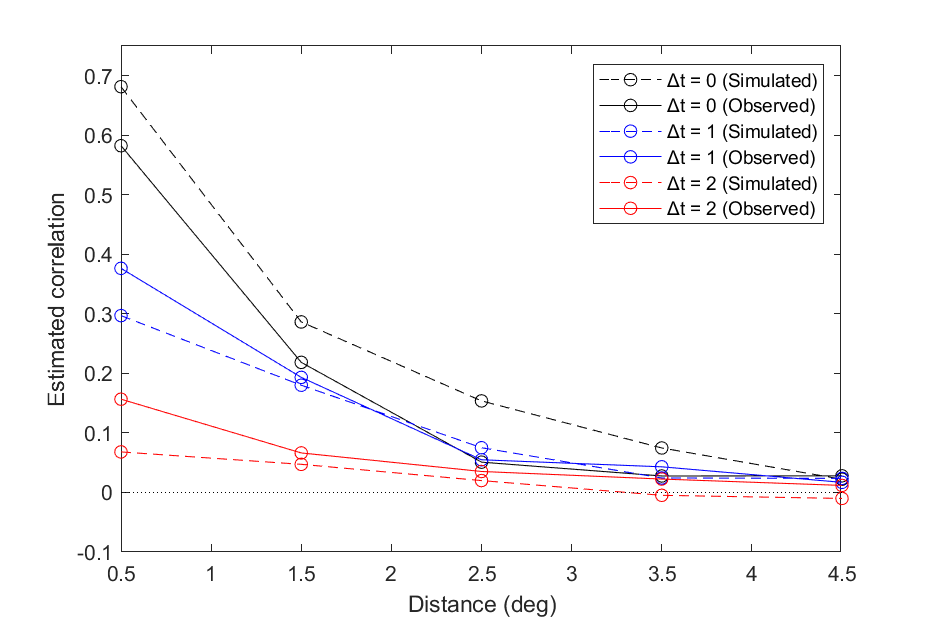}}
    \subfigure[Midocean (univariate)]{
        \includegraphics[width=7.5cm,trim = 0.5cm 0cm 1cm 0.5cm,clip = true]{./bivariate_figures/intTempCVPredsLocalCondSimCorrBivariateLOFOSpaceTimeTrend15_975_2005_2005_2_all.png}}
    \subfigure[Midocean (bivariate)]{
       \includegraphics[width=7.5cm,trim = 0.5cm 0cm 1cm 0.5cm,clip = true]{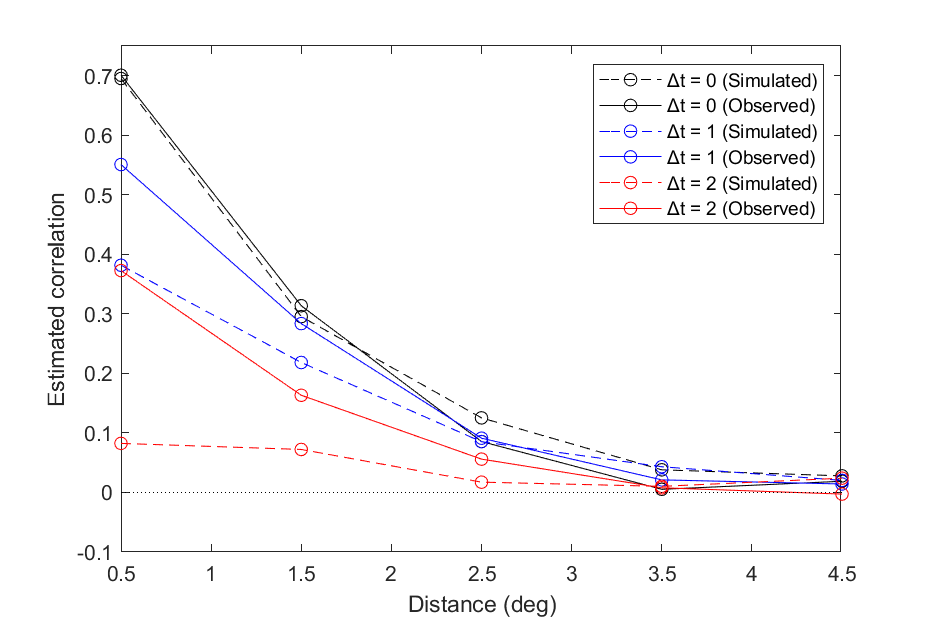}}
       \subfigure[Cross-layer]{
       \includegraphics[width=7.5cm,trim = 0.5cm 0cm 1cm 0.5cm,clip = true]{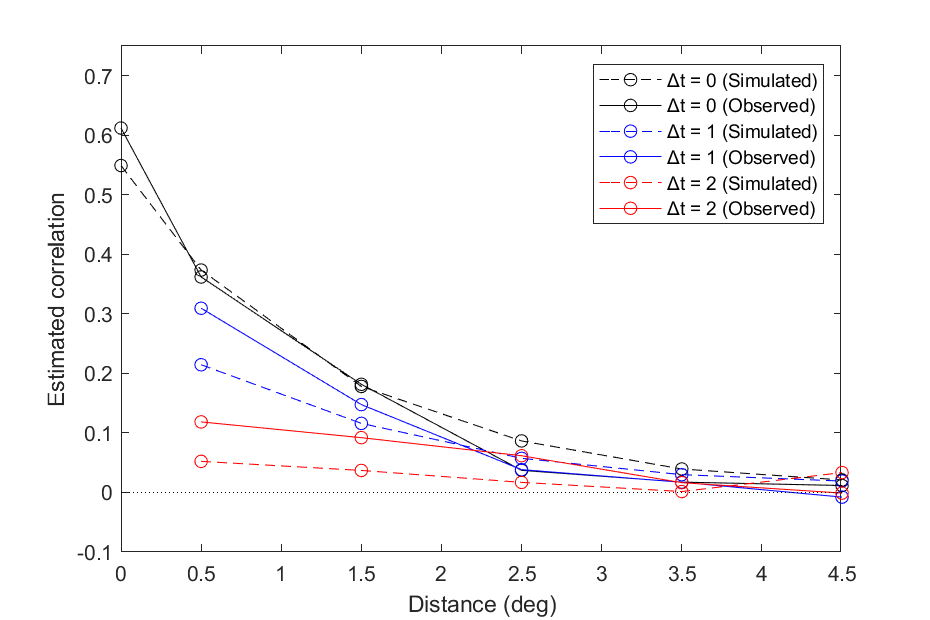}}
    \caption{Cross-validation for conditional spatio-temporal dependence. The solid and dashed lines represent the observed and conditionally simulated empirical correlation respectively.}
    \label{fig:cv_condsim}
\end{figure}

We randomly subsample up to 120,000 (due to computational limitations) vertically integrated temperature profile pairs in 2005 for each subplot to produce Figure \ref{fig:cv_condsim}, where the temporal lag $\Delta t$ ranges from 0 to 2 months and the spatial lag $\Delta d$ 0.5 to 4.5 degrees in 1-degree bins (the cross-layer spatial lag ranges from 0 to 4.5). We notice that in the upper ocean, the bivariate model has comparable performance to the univariate model in capturing the observed conditional dependence. The bivariate model also appears to broadly capture the observed cross-layer conditional dependence. However, in \cite{univariate}, the model seemed to underestimate the observed conditional dependence at longer time lags (red lines) in the midocean, which also persists for the bivariate model. This could indicate that there is room for improvement in modeling the covariance structure—this is a topic for future investigation and we provide a brief discussion in Section~\ref{sec:discussion}.

\subsection{Ocean heat content uncertainties}

\begin{figure}[!h]
\centering\noindent
\includegraphics[width=8cm,trim = 1.5cm 1cm 0.8cm 2cm,angle=0]{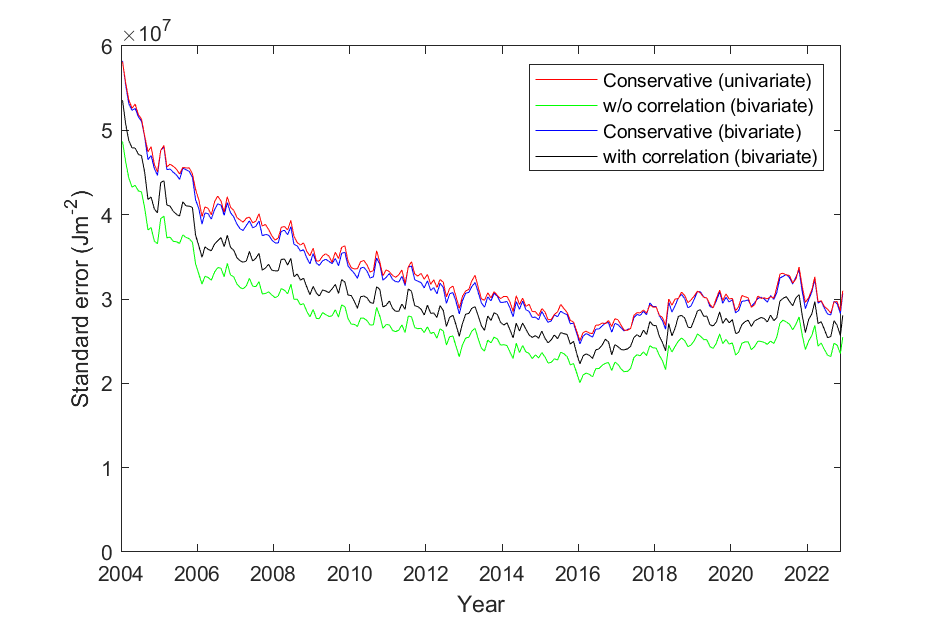}\\
  \caption{Total global OHC anomaly (15--1850 dbar) standard error comparison. The red and green lines show baseline uncertainty estimates, while the blue, black, and green lines show the bivariate model uncertainty estimates.}\label{fig:stderr_diff}
\end{figure}

We plot the monthly total global OHC (Eq. \ref{eq:OHC_def_xy}) anomaly standard errors in Figure \ref{fig:stderr_diff} and show the
univariate conservative upper bound (Eq. \ref{eq:conservative-bound}; red) and the bivariate model uncertainty estimates ignoring the vertical correlation (i.e. ignoring the conditional covariance term in Equation \eqref{eq:cond-var-total}, which underestimates the uncertainty; green) as baselines for comparison. We see that evaluating the conservative upper bound using conditional simulation realizations from the bivariate model results
in a slight reduction in the uncertainties (blue), which is supported by the overall reduction in predictive
variance (Figure \ref{fig:var_diff}). Since the bivariate model incorporates the vertical correlation between the upper ocean and midocean, we are able to produce an improved uncertainty estimate (black). We
see that these uncertainties are consistently up to 15\% reduced from the univariate conservative upper bound. In Section \ref{sec:ohc-estimates}, we demonstrate that this reduction is also evident in the OHC trend and OHU (ocean heat uptake) uncertainties.

\subsection{Statistical significance}

\begin{figure}[!h]
\centering
\subfigure[Significant points with respect to time]{
        \includegraphics[width=7cm,trim = 0.5cm 0cm 1cm 0.5cm,clip = true]{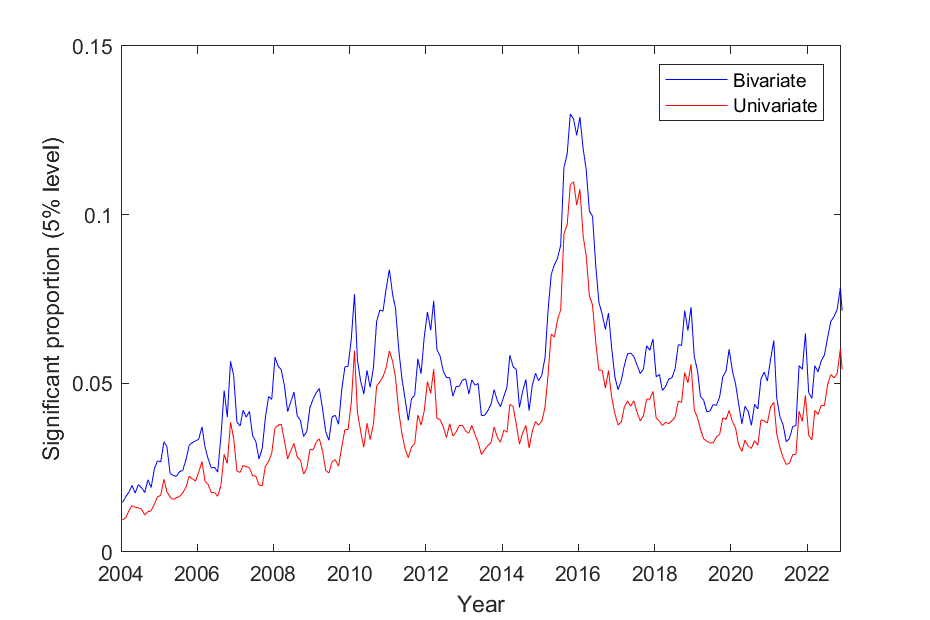}}
    \subfigure[Significant points with respect to space]{
        \includegraphics[width=8cm,trim = 0.5cm 0.5cm 1cm 0.5cm,clip = true]{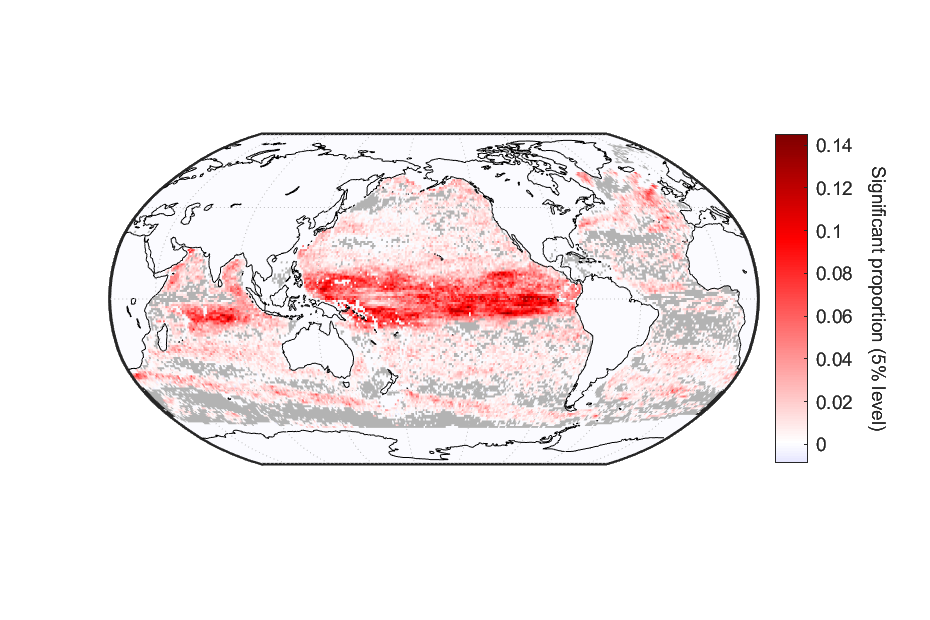}}
      \subfigure[Regional OHC trends]{\includegraphics[width=8cm,trim = 0cm 2cm 0cm 0cm,clip = true]{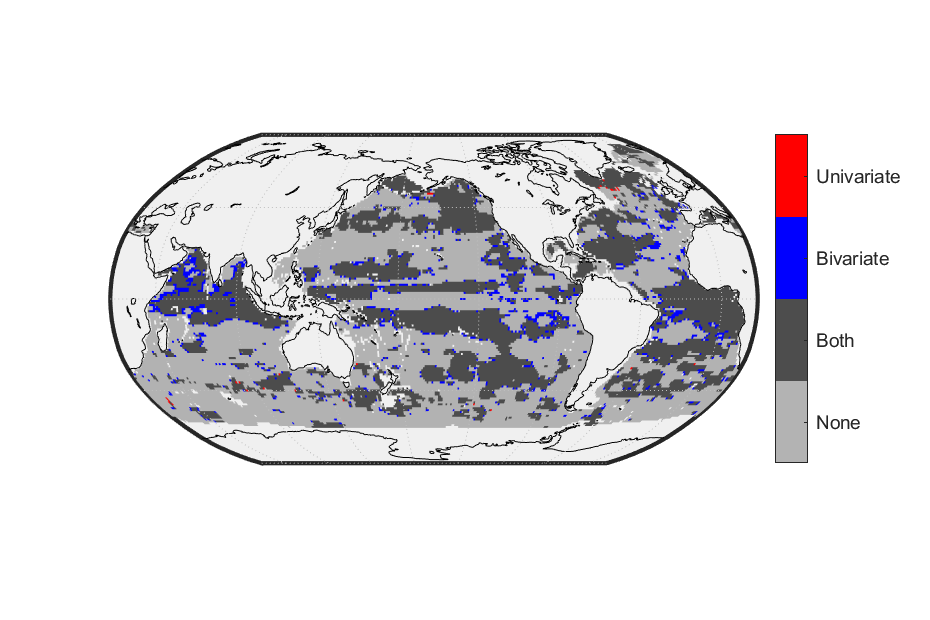}}
  \caption{Comparison of statistical significance for total regional OHC anomalies (15-1850 dbar; 5\% level). Figure (a): Monthly proportion of significant grid points across the Argo-sampled global ocean. Figure (b): Proportion of significant months by grid point. The colorbar shows the proportion difference compared to the univariate model. Figure (c): Univariate and bivariate model regional trends with significant trends (5\% level) highlighted.}\label{fig:sigprop_diff}
\end{figure}

Furthermore, we investigate the effect of modeling vertical dependence on statistical significance at regional scales. The left panel of Figure \ref{fig:sigprop_diff} shows the monthly proportion of grid points with statistically significant OHC anomalies (5\% level) across the Argo-sampled part of the global ocean, where the critical points for each grid point are determined by the grid point-wise standard error of the total OHC anomaly conditional simulation realizations. Similar to above, we observe that the bivariate model (blue) consistently produces a larger proportion of significant grid points over time compared to the univariate (red). In other words, the bivariate model is able to more efficiently extract the transient regional OHC signal from the Argo data. The large spike in significant grid points around 2016 likely corresponds to the strong El Ni\~{n}o-Southern Oscillation (ENSO) phenomenon in late 2015-early 2016. In Section \ref{sec:ohc-estimates}, we show that we can also use the bivariate model to improve uncertainties on the cross-correlation between ENSO and OHC. The right panel of Figure \ref{fig:sigprop_diff} shows the difference between the in the proportion of significant months for each grid point between 2004 and 2022. We can see that in the vast majority of regions, modeling the vertical dependence results in an increase in the proportion of significant months (red). This increase appears to be most prominent in the equatorial region, which may be explained by the positive estimated nugget correlation in the equatorial region (shown in the appendix). Finally, in the bottom panel, we plot the grid points with significant OHC trends at the 5\% level. We can see that the regions where the OHC trend is significant only under the bivariate model (blue) provide a noticeable expansion to the regions where the trend is significant under both models (dark gray).

\section{Example applications for bivariate ocean heat content estimates with uncertainties}
\label{sec:results}
In this section, we present the estimated cross-correlation parameters and a number of case studies, which highlight the bivariate model's improved uncertainties on downstream oceanographic quantities. 

\subsection{Estimated cross-correlation parameters}
\begin{figure}[!h]
  \centering\noindent\includegraphics[width=8cm,trim = 1.5cm 3.5cm 0.8cm 3cm,angle=0,clip=true]{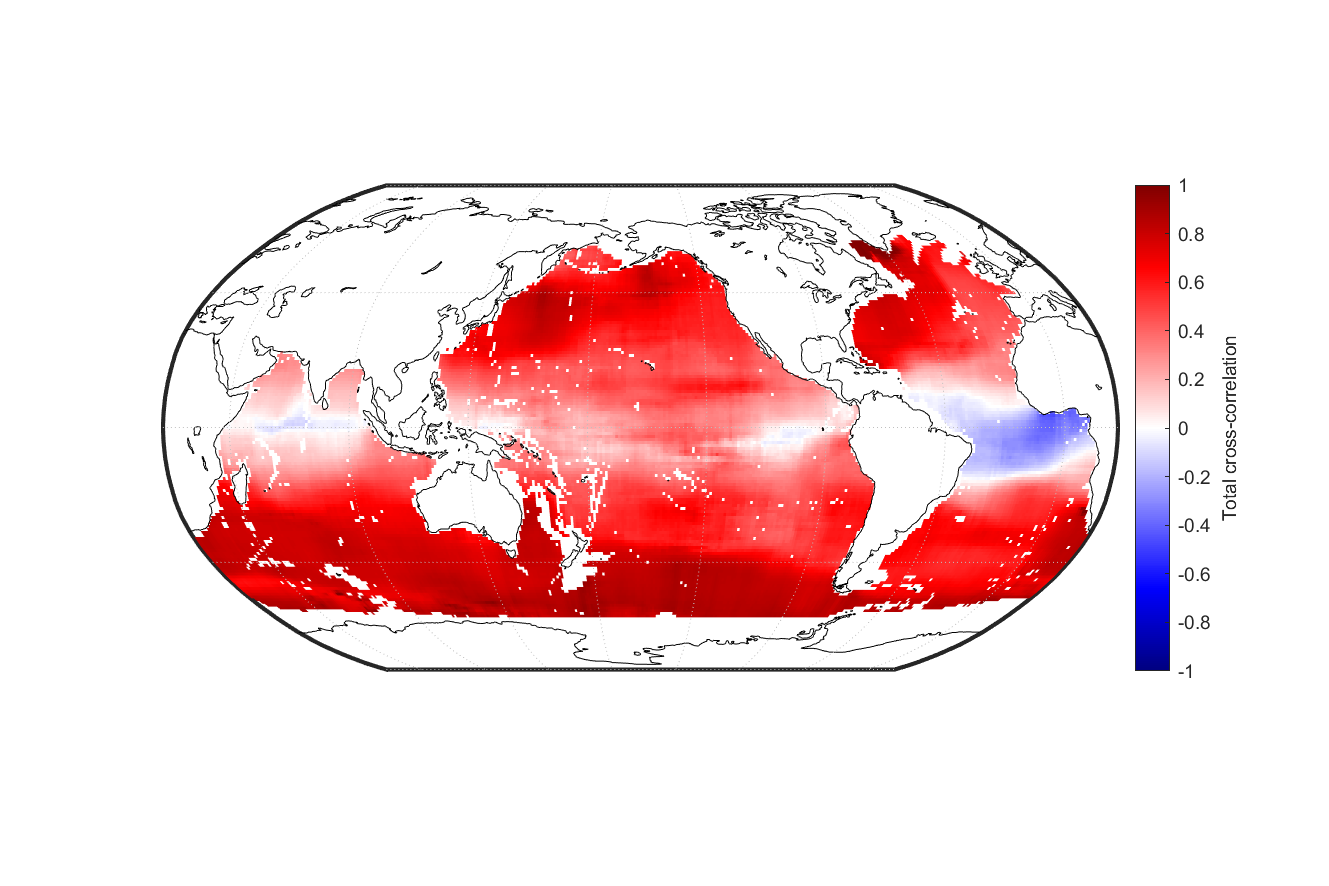}
  \caption{Estimated total vertical cross-correlation between 15--975 dbar and 975--1850~dbar OHC.}\label{fig:total-corr}
\end{figure}

Figure \ref{fig:total-corr} shows the estimated total cross-correlation $\widehat{\text{Corr}}(\widetilde{\OHC_1},\widetilde{\OHC_2})=a\hat \beta + b\hat\rho$ where 
\begin{equation}
a = \frac{\sqrt{\hat\phi_1}\sqrt{\hat\phi_2}}{\sqrt{\hat\phi_{1}+\hat\sigma_1^2}+\sqrt{\hat\phi_{2}+\hat\sigma_2^2}} \quad 
b = \frac{\hat\sigma_1\hat\sigma_2}{\sqrt{\hat\phi_{1}+\hat\sigma_1^2}+\sqrt{\hat\phi_{2}+\hat\sigma_2^2}}
\end{equation}
We provide the derivation for the coefficients $a$ and $b$ in Section \ref{sec:crosscorr-coef} in the appendix. For brevity, we also refer the reader to the appendix for the individual plots of the estimated field cross-correlation $\hat \beta$ and nugget cross-correlation $\hat \rho$. In most regions, we find that the the estimated total cross-correlation between the upper and midocean OHC residuals is strong and positive. We estimate the cross-correlation to be particularly strong and positive near the major ocean currents (the Kuroshio Current in the Northwestern Pacific, the Gulf Stream in the Northwestern Atlantic, and the Agulhas Current in the Southwestern Indian Ocean). This is likely because meanders and eddies in these currents extend deep into the water column--these regions indicate where we expect to see the greatest improvement from modeling the vertical dependence. Note that we estimate the cross-correlation to be negative in the equatorial Atlantic, which is consistent with scatterplots of the vertically integrated OHC residuals 
 (Figure \ref{fig:atlantic-scatterplots} in the appendix). We also see in Figures \ref{fig:blob-scatterplots} (northeast Pacific) and \ref{fig:indianocean-scatterplots} (southern Indian Ocean) in the appendix that the empirical correlation is strong and positive in these regions which is consistent with our model-based estimates in Figure~\ref{fig:total-corr}. Therefore, the sign and strength of the estimated correlation seem to match scatterplots of the data. This highlights the bivariate model's explanatory potential on regional ocean dynamics.

\label{sec:param-estimates}

\subsection{Bivariate global OHC trend}
\label{sec:ohc-estimates}
One of the main advantages of the local conditional simulation approach to uncertainty quantification is flexibility. Throughout Sections 3 and 4, we emphasize improving uncertainties for OHC, but as noted in \cite{univariate}, it is also possible to obtain uncertainties for transformations of OHC by simply applying the relevant transformation to the conditional simulation realizations. For example, to obtain an uncertainty on the total OHC trend, we first fit a linear regression model with seasonal factors to each total OHC conditional simulation realization time series using ordinary least-squares. We then obtain the standard error of the trend by computing the sample standard deviation of the ensemble of estimated trends. 

\begin{figure}[!h]
    \centering
    \includegraphics[width=7.5cm,trim = 0.5cm 0.1cm 1cm 0cm,clip = true]{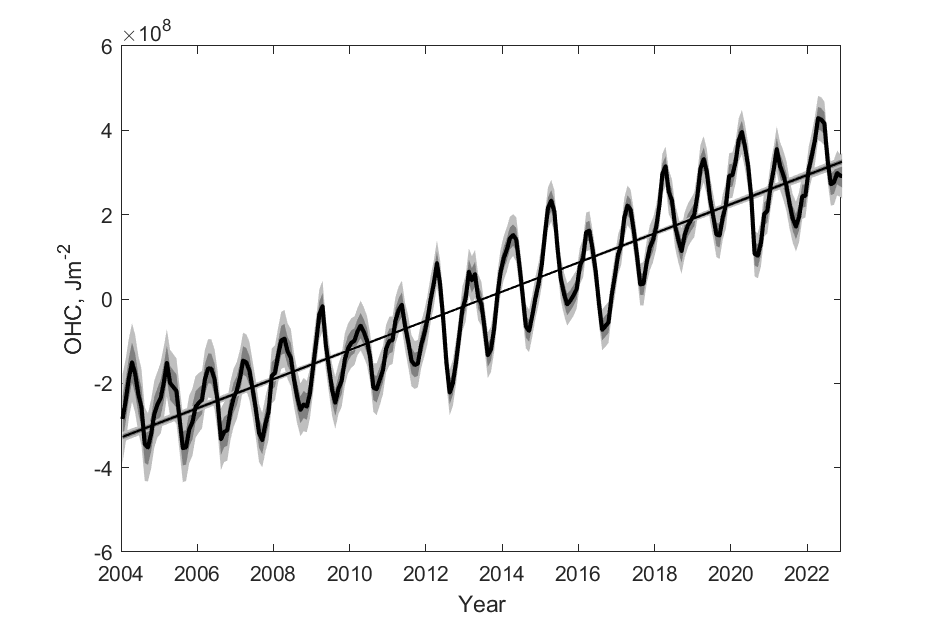}
\caption{Total OHC (15--1850 dbar) time series with 68\% (dark gray) and 95\% (light gray) uncertainties. The trend estimate with 95\% uncertainty is $1.115$ W/m$^2$ $\pm$ $0.020$ W/m$^2$.}
    \label{fig:trend_bivariate}
\end{figure}

We observe that the estimated global OHC trend from the bivariate model (Figure \ref{fig:trend_bivariate}; left) is comparable to the univariate estimate of 1.114 W/m$^2$, but the $95\%$ uncertainty when we incorporate the vertical dependence is
reduced by approximately 13\% from $0.024$ to $0.020$ W/m$^2$. 

\subsection{Ocean heat uptake}
\begin{figure}[!h]
    \centering
    \subfigure[Monthly]{
        \includegraphics[width=6cm,trim = 0.5cm 0.1cm 1cm 0.5cm,clip = true]{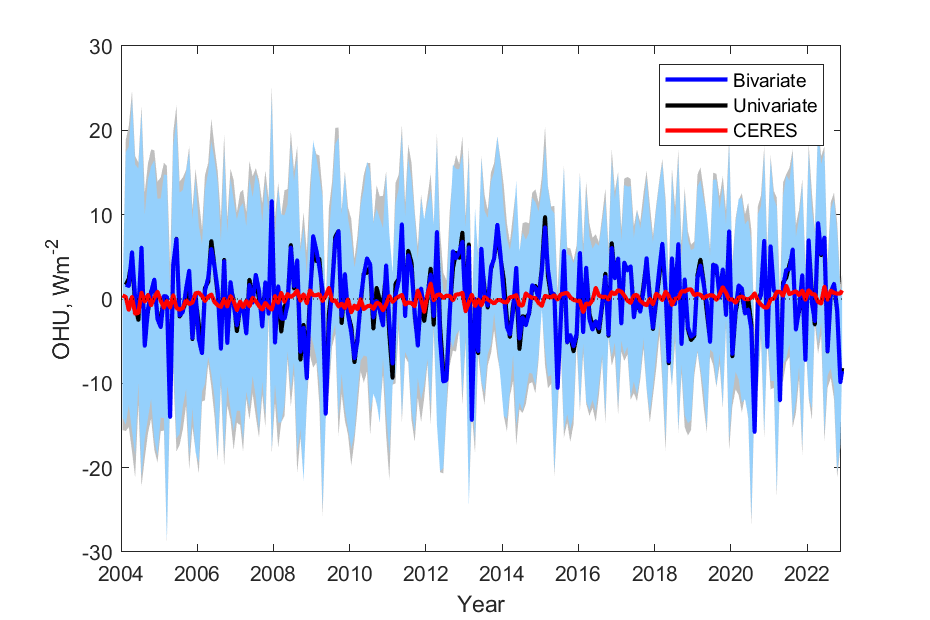}}
    \subfigure[12-month moving average]{
        \includegraphics[width=6cm,trim = 0.5cm 0.1cm 1cm 0.5cm,clip = true]{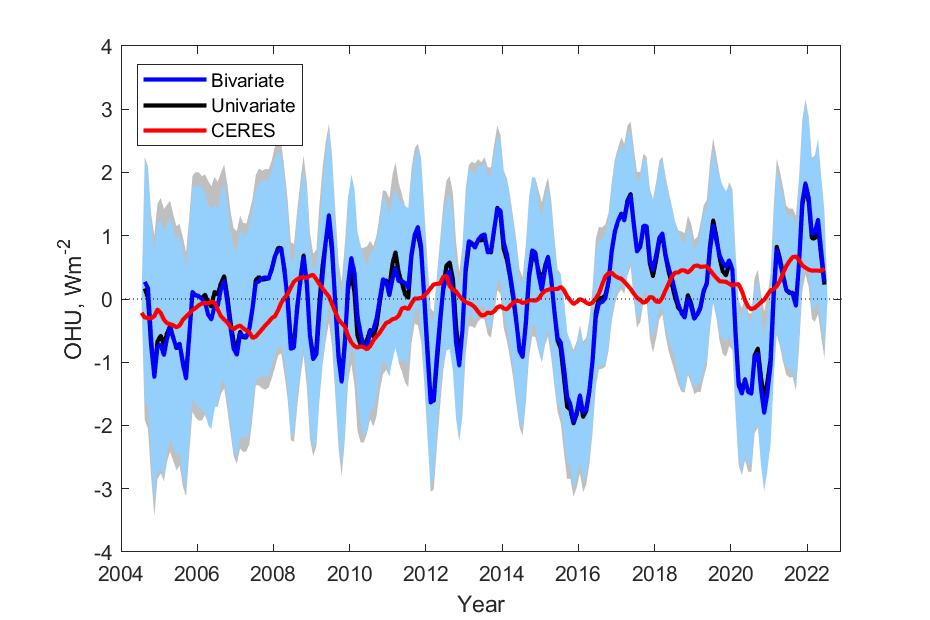}}\\
    \subfigure[24-month moving average]{
        \includegraphics[width=6cm,trim = 0.5cm 0.1cm 1cm 0.5cm,clip = true]{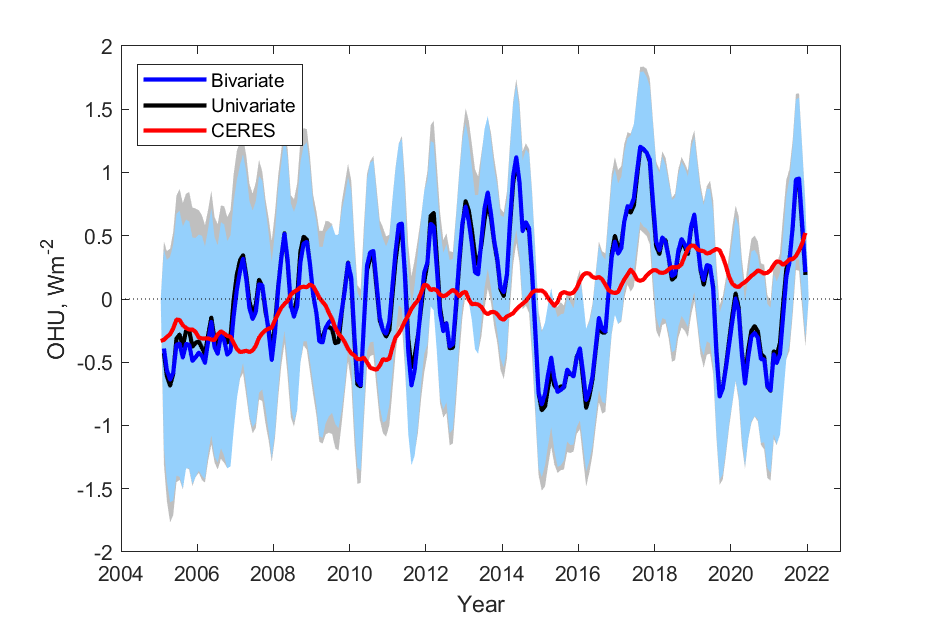}}
    \subfigure[36-month moving average]{
        \includegraphics[width=6cm,trim = 0.5cm 0.1cm 1cm 0.5cm,clip = true]{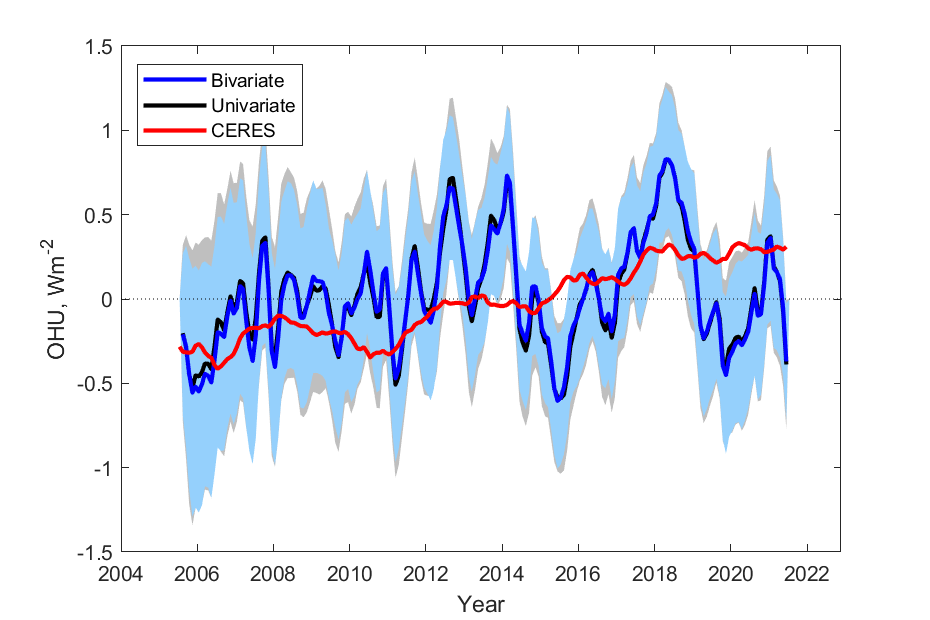}}\\
    \caption{Ocean heat uptake (15--1850 dbar) with 95\% uncertainties (gray = univariate conservative, blue = bivariate) compared with CERES top-of-atmosphere radiative net flux anomalies.}
    \label{fig:ohu}
\end{figure}

We next consider the ocean heat uptake (OHU) uncertainties. OHU is defined as the first derivative of OHC with respect to time (i.e. $\text{OHU}(t) = \mathrm{d}/\mathrm{d}t \; \text{OHC}(t)$). Comparing OHU and top-of-atmosphere (TOA) energy fluxes is often of interest in climate science because radiation imbalance at the top of the atmosphere influences interannual variability and long-term changes in OHC. In Figure \ref{fig:ohu}, we plot the univariate (black) and bivariate (blue) OHU estimates alongside CERES satellite TOA net flux measurements (red) from the EBAF-TOA 4.2 data product (e.g. \cite{Loeb2018}, \cite{Kato2018}). Notice that since the bivariate model is able to avoid the conservative upper bound, a reduction in uncertainty is also present for the OHU estimates - the differences between the univariate (light gray) and bivariate (light blue) 95\% confidence intervals tend to be most prominent in the earlier years of the Argo sampling period (around 2004--2008). This suggests that the additional information incorporated into the bivariate model may be especially beneficial during sampling periods when observations are sparse.   

\subsection{Cross-correlation of OHC anomalies with ENSO}
\begin{figure}[!h]
  \centering\noindent\includegraphics[width=8cm,angle=0]{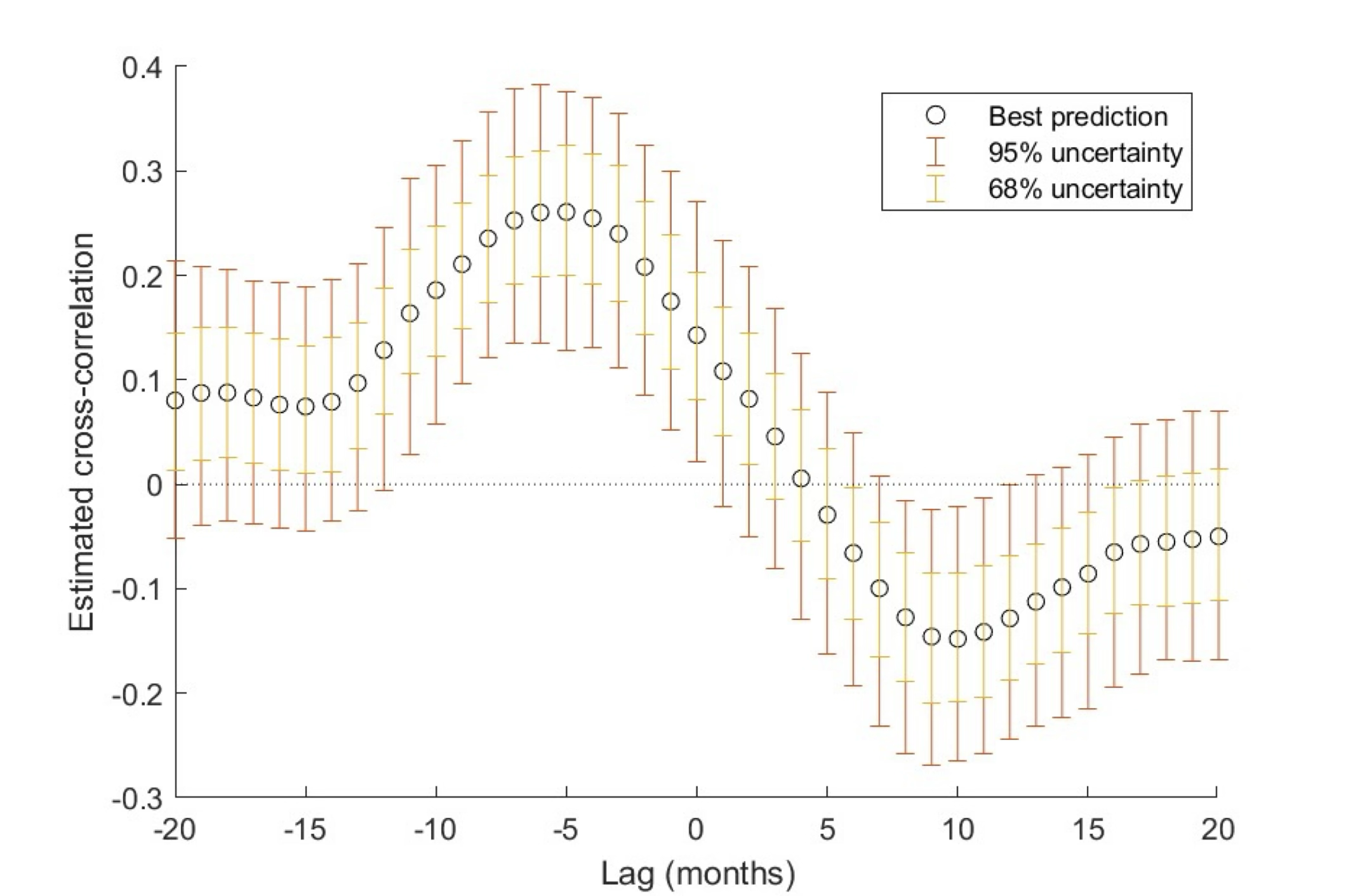}\\

    \caption{Cross-correlation of monthly global OHC anomalies (15--1850 dbar) with 68\% and 95\% uncertainties vs monthly input SST anomalies to the ONI index.}\label{fig:oni_xcorr}
\end{figure}

Finally, in Figure \ref{fig:oni_xcorr}, we present uncertainties on the cross-correlation between the total OHC anomalies and the monthly sea surface temperature (SST) input to the NOAA Oceanic Nino Index (ONI) \citep{ONI}. The ONI index is a 3-month moving average of SST anomalies in the Niño 3.4 region ($5^\circ$N--$5^\circ$S, $120^\circ$W--$170^\circ$W). It is generally used to classify whether the ENSO phenomenon is present: if the ONI is at least $0.5^\circ$C for 5 consecutive months, the ocean is in El Niño state (conversely, if the ONI is $\leq -0.5^\circ$C for at least 5 consecutive months, then the ocean is in La Niña state). Monitoring ENSO is important because of its consequences on global weather patterns (e.g., \cite{Trenberth2020}). Note that the conservative bound on the total OHC uncertainty in Equation \eqref{eq:conservative-bound} does not easily generalize to the cross-correlation between OHC anomalies and monthly SST. For this reason, \cite{univariate} provides limited uncertainty quantification for the upper ocean cross-correlation only. Since we now incorporate the vertical dependence between the upper ocean and midocean in the bivariate estimates, full uncertainty quantification is enabled for the water column between 15--1850 dbar. Although the total OHC anomaly estimates are comparable to the univariate estimates and also seem to lead the monthly SST by 6 months (the magnitude of the estimated cross-correlation is slightly stronger; +0.05), the bivariate model enables more complete inference by including the midocean contribution.

\section{Discussion}
\label{sec:discussion}

In this paper, we presented an improved OHC mapping and uncertainty quantification framework that incorporates vertical dependence between pressure layers. Based on \cite{univariate}, we developed a bivariate extension of the locally stationary Gaussian process regression model that strengthens predictive performance by enabling sparsely sampled regions deeper in the water column to borrow information from well-sampled regions closer to the surface. In addition, we demonstrated that modeling the vertical dependence results in substantial reductions in OHC anomaly uncertainties and many downstream oceanographic quantities of interest. Through this framework, we were able to achieve statistically rigorous, computationally feasible uncertainties for the total OHC (15--1850 dbar) from regional to global scales. This work motivates many directions for further research in terms of both statistical method development and scientific applications.

It is important to note that although the use of local methods made parameter estimation and conditional simulation computationally feasible at the scale of Argo data, we would require additional computational improvements to enable various further extensions. These tasks remain computationally intensive due to slow matrix operations. For example, obtaining 500 conditional simulation realizations for a single month from the bivariate model required approximately 2.5 hours on the Bridges-2 supercomputer. One way to reduce computation time is to apply the neural likelihood or analogous neural conditional simulation machinery in \cite{Walchessen2024} and \cite{Walchessen2025} which carries the benefit of amortization \citep{ZammitMangion2025}. This would, however, be a non-trivial extension as Argo observations do not lie on a regular grid; we would, for example, need to replace the convolutional neural network (CNN) in the above works with an architecture capable of handling irregular inputs, such as a graph neural network (GNN) as in \cite{Sainsbury-Dale2025}.

These computational improvements would eventually enable incorporating additional information such as satellite measurements (e.g., sea surface temperature and sea surface height) into the modeling process, which we expect to increase the predictive accuracy of the mapped anomalies as demonstrated in \cite{LymanJohnson23}. Jointly modeling more than two fields would likely also become feasible—potential directions to consider are to map the two OHC layers and SST, or to split the vertical column into more than two pressure layers. In addition to improved maps and reduced uncertainty, the estimated cross-correlation parameters may reveal scientifically valuable insights on the relationship between these variables.

Our results in Section \ref{sec:validation} also motivate a number of statistical improvements to the model definition. For example, the cross-validation plots in Section \ref{sec:cv-condsim} suggest that there may be a slight misspecification in the spatio-temporal dependence model. There is a large body of literature (e.g., \cite{Porcu2021}) spanning more than thirty years of work on space-time covariance models that could provide insight on more flexible models to better capture the dependence structure, especially at longer time scales.  

Furthermore, based on scatterplots of the observations in certain ocean regions (e.g., the Kuroshio region) we investigated early in this work, it appears there are patterns that are not characteristic of the elliptical shapes we would expect from a bivariate Gaussian distribution. The cross-validation study in \cite{Kuusela2018} also revealed heavy-tailed observations. In addition to considering more flexible covariance functions, accounting for this non-Gaussian structure would likely improve the maps and their uncertainties. Two potential ways to achieve this are to consider non-Gaussian convolution models (\cite{Higdon2002}) or the SPDE approach (\cite{Bolin2019}) mentioned in Section \ref{sec:background}.

\clearpage
\section*{Acknowledgments}

TS and MK were supported by NOAA grant NA21OAR4310258. DG was supported by NOAA grant NA21OAR4310261. We would like to thank members of the CMU STAMPS Research Center as well as participants of the GEWEX Earth Energy Imbalance Assessment Workshops for helpful comments and feedback on this work. This work used Bridges-2 at Pittsburgh Supercomputing Center through allocation MTH230014 from the Advanced Cyberinfrastructure Coordination Ecosystem: Services \& Support (ACCESS) program, which is supported by National Science Foundation grants \#2138259, \#2138286, \#2138307, \#2137603, and \#2138296.

\newpage

\appendix
\section{Appendix}
\subsection{Estimated cross-correlation parameters}

\begin{figure}[!h]
    \centering
    \subfigure[Field cross-correlation $\beta$]{
        \includegraphics[width=8cm,trim = 1.5cm 3cm 0.8cm 2cm,clip = true]{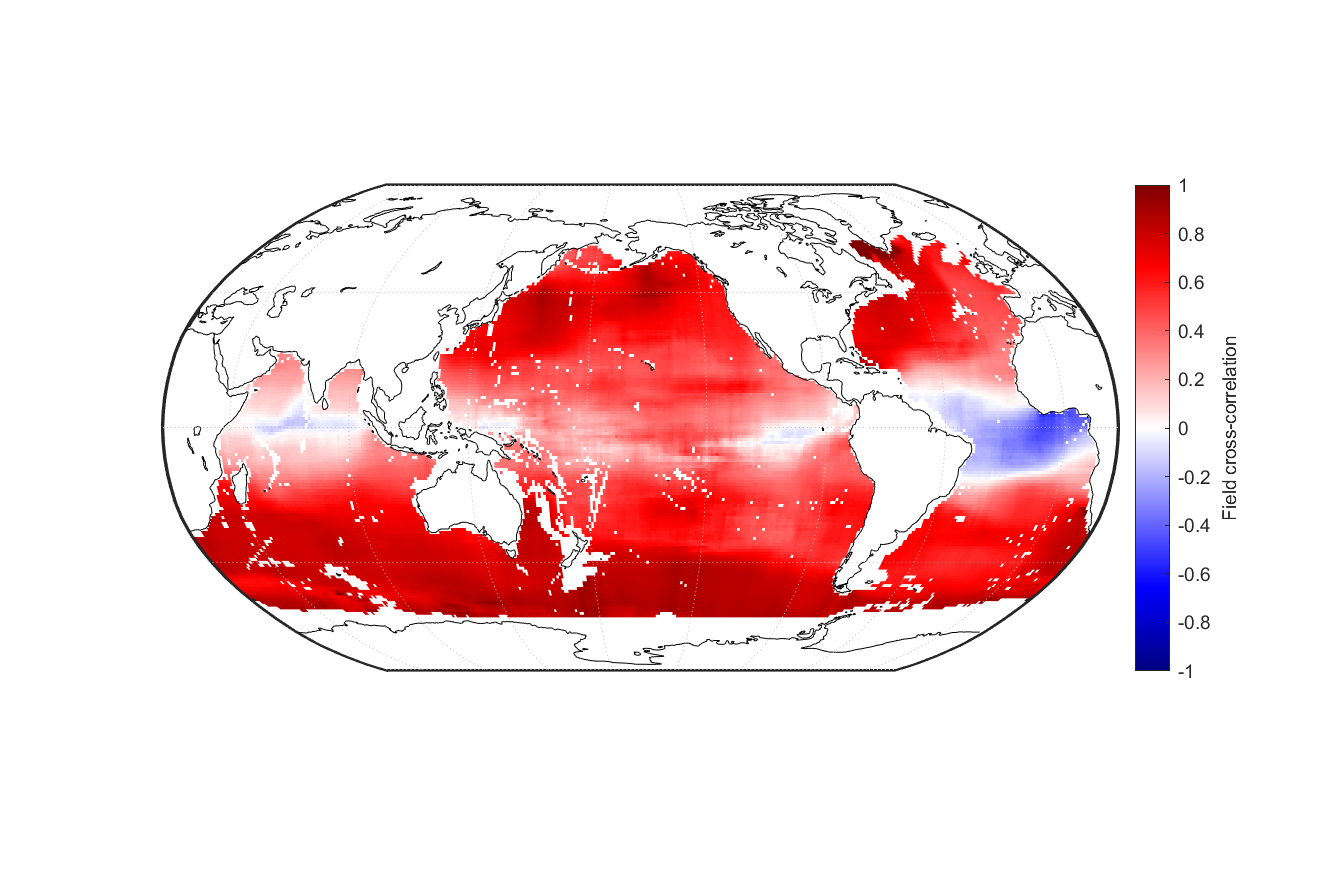}}
    \subfigure[Nugget cross-correlation $\rho$]{
        \includegraphics[width=8cm,trim = 1.5cm 3cm 0.8cm 2cm,clip = true]{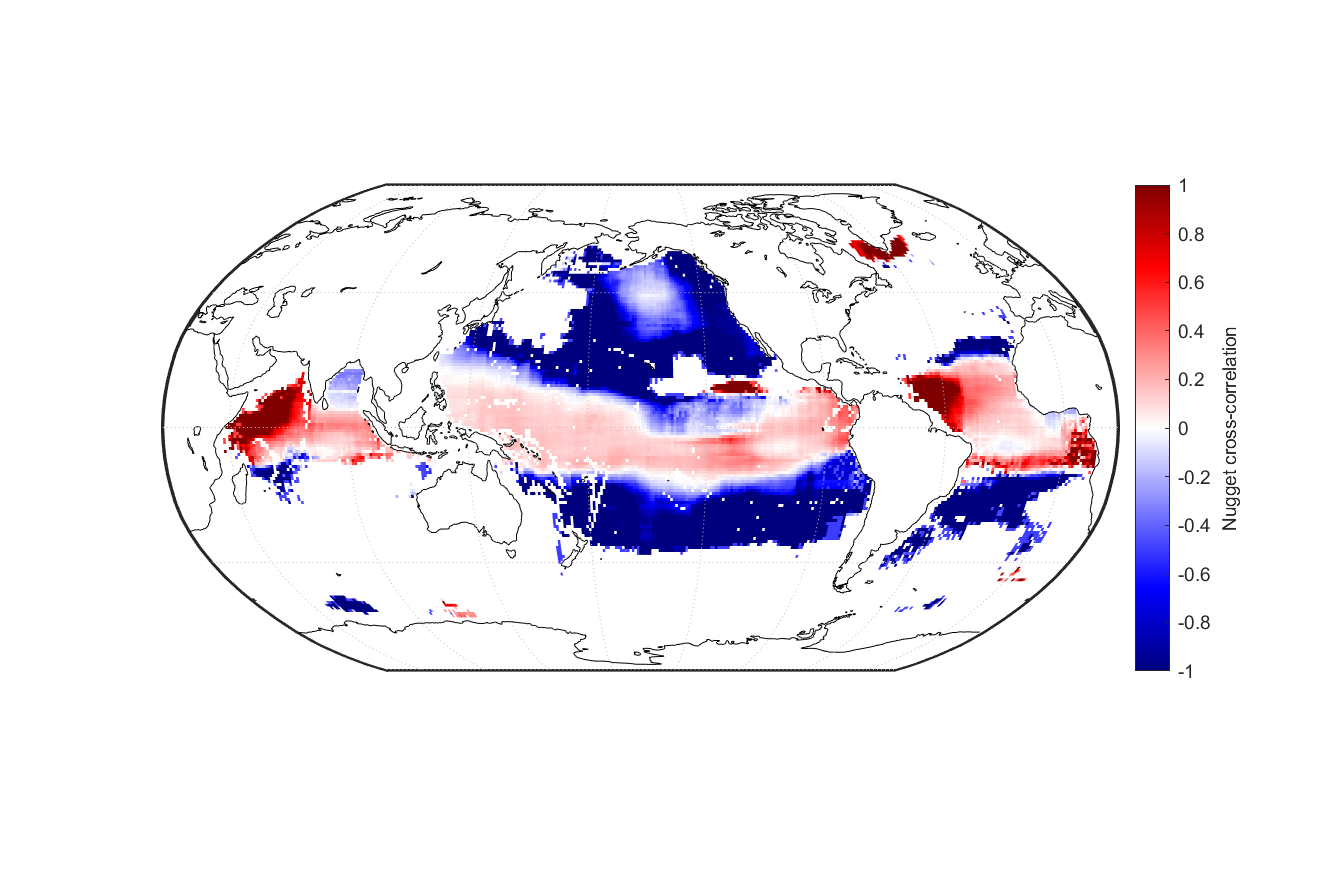}}
    \caption{Estimated cross-correlation parameters}
    \label{fig:cross-correlation-appendix}
\end{figure}

\clearpage
\subsection{Total cross-correlation derivation}
\label{sec:crosscorr-coef}
As in Section \ref{sec:local-gp}, let $a_l(\boldsymbol{z})$ be the anomaly field and $\varepsilon_l(\boldsymbol{z})$ be the nugget effect for pressure layer $l$ where $\boldsymbol{z}=(x,y,t)$. Then the correlation between the mean-subtracted OHC residuals in pressure layers $1$ and $2$ is
\begin{align*}
\text{Corr}(a_1(\boldsymbol{z})+\varepsilon_1(\boldsymbol{z}),a_2(\boldsymbol{z})+\varepsilon_2(\boldsymbol{z})) &=\\ \frac{\text{Cov}(a_1(\boldsymbol{z}), a_2(\boldsymbol{z}))+\text{Cov}(a_1(\boldsymbol{z}),\varepsilon_2(\boldsymbol{z}))+\text{Cov}(\varepsilon_1(\boldsymbol{z}),a_2(\boldsymbol{z}))+\text{Cov}(\varepsilon_1(\boldsymbol{z}),\varepsilon_2(\boldsymbol{z}))}{\sqrt{(\text{Var}(a_1(\boldsymbol{z}))+2\text{Cov}(a_1(\boldsymbol{z}),\varepsilon_1(\boldsymbol{z}))+\text{Var}(\varepsilon_1))(\text{Var}(a_2(\boldsymbol{z}))+2\text{Cov}(a_2(\boldsymbol{z}),\varepsilon_2(\boldsymbol{z}))+\text{Var}(\varepsilon_2))}}) &=\\
\frac{\beta \sqrt{\phi_1}\sqrt{\phi_2}+\rho \sigma_1\sigma_2}{\sqrt{(\phi_1+\sigma_1^2)(\phi_2+\sigma_2^2)}}.
\end{align*}
This implies that
\begin{align*}
\widehat{\text{Corr}}(\widetilde{\OHC_1},\widetilde{\OHC_2}) &= \frac{\sqrt{\hat\phi_1}\sqrt{\hat\phi_2}}{\sqrt{\hat\phi_{1}+\hat\sigma_1^2}+\sqrt{\hat\phi_{2}+\hat\sigma_2^2}}\hat \beta
+ \frac{\hat\sigma_1\hat\sigma_2}{\sqrt{\hat\phi_{1}+\hat\sigma_1^2}+\sqrt{\hat\phi_{2}+\hat\sigma_2^2}}\hat \rho 
\\&:=a\hat \beta + b\hat\rho.
\end{align*}

\clearpage
\subsection{Regional scatterplots}

\begin{figure}[!h]
    \centering
        \includegraphics[width=8cm,trim = 0.5cm 0cm 1cm 0cm,clip = true]{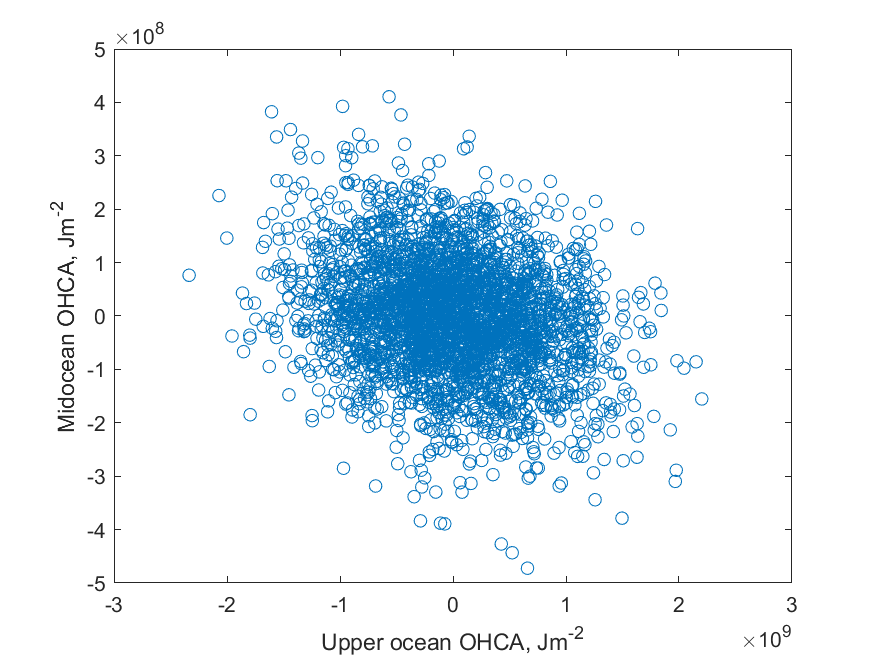}
    \caption{Vertically integrated OHC residuals for the collection of $20^\circ \times 20^\circ \times 3$ month windows centered on February 2004--2022, shown for a location in the equatorial Atlantic Ocean ($3.5^\circ$S, $10.5^\circ$E). The empirical Pearson correlation is -0.2455.}
    \label{fig:atlantic-scatterplots}
\end{figure}

\begin{figure}[!h]
    \centering
    
        \includegraphics[width=8cm,trim = 0.5cm 0cm 1cm 0cm,clip = true]{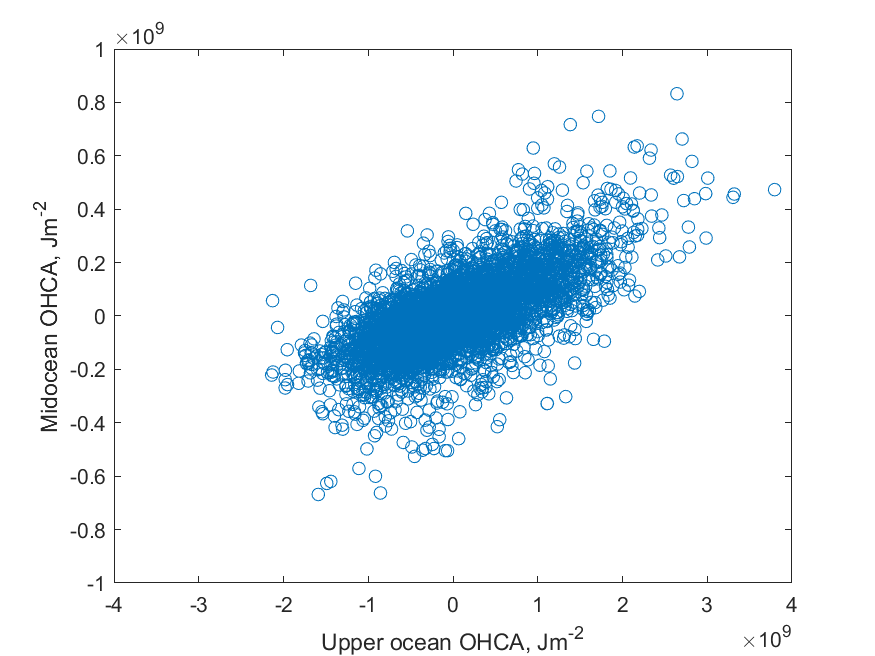}
    
    \caption{Vertically integrated OHC residuals for the collection of $20^\circ \times 20^\circ \times 3$ month windows centered on February 2004--2022, shown for a location in the the northeast Pacific Ocean ($48.5^\circ$N, $141.5^\circ$W). The empirical Pearson correlation is 0.6647.}
    \label{fig:blob-scatterplots}
\end{figure}

\begin{figure}[!h]
    \centering
        \includegraphics[width=8cm,trim = 0.5cm 0cm 1cm 0cm,clip = true]{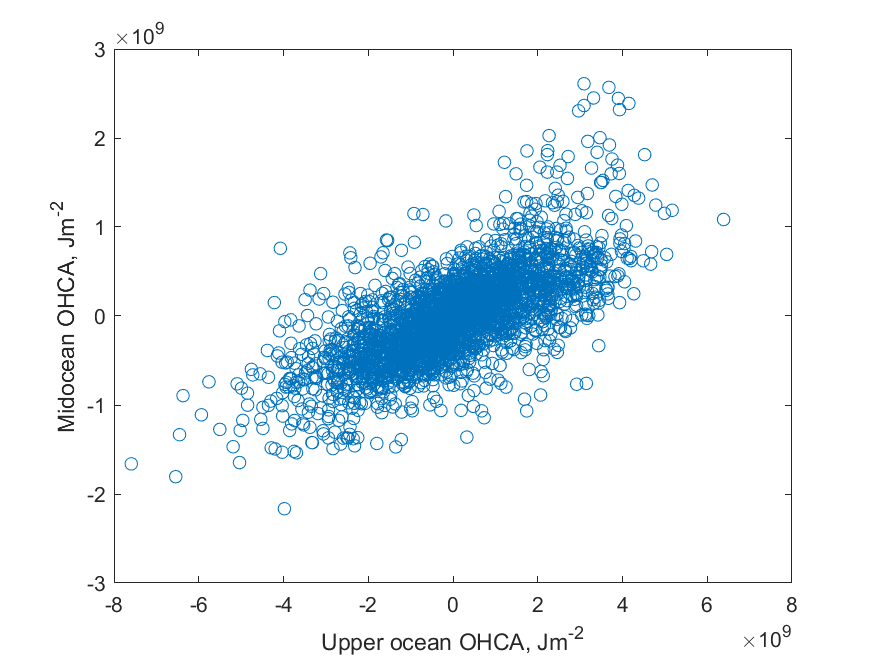}
    \caption{Vertically integrated OHC residuals for the collection of $20^\circ \times 20^\circ \times 3$ month windows centered on February 2004--2022, shown for a location in the southern Indian Ocean ($29.5^\circ$S, $69.5^\circ$E). The empirical Pearson correlation is 0.7024.}
    \label{fig:indianocean-scatterplots}
\end{figure}

\clearpage
\bibliographystyle{agsm}
\bibliography{refs}

@article{Walchessen2024,
title = {Neural likelihood surfaces for spatial processes with computationally intensive or intractable likelihoods},
journal = {Spatial Statistics},
volume = {62},
pages = {100848},
year = {2024},
issn = {2211-6753},
doi = {https://doi.org/10.1016/j.spasta.2024.100848},
author = {Julia Walchessen and Amanda Lenzi and Mikael Kuusela}
}

@article{Sainsbury-Dale2025,
author = {Matthew Sainsbury-Dale and Andrew Zammit-Mangion and Jordan Richards and Raphaël Huser},
title = {Neural {Bayes} Estimators for Irregular Spatial Data Using Graph Neural Networks},
journal = {Journal of Computational and Graphical Statistics},
volume = {34},
number = {3},
pages = {1153–1168},
year = {2025},
publisher = {ASA Website},
doi = {10.1080/10618600.2024.2433671}
}

@misc{Walchessen2025,
      title={Neural Conditional Simulation for Complex Spatial Processes}, 
      author={Julia Walchessen and Andrew Zammit-Mangion and Raphaël Huser and Mikael Kuusela},
      year={2025},
      howpublished={arXiv:2508.20067 [stat.ME]}
}

@article{Porcu2021,
author = {Porcu, Emilio and Furrer, Reinhard and Nychka, Douglas},
title = {30 Years of space–time covariance functions},
journal = {WIREs Computational Statistics},
volume = {13},
number = {2},
pages = {e1512},
doi = {10.1002/wics.1512},
year = {2021}
}

@InProceedings{Higdon2002,
author="Higdon, Dave",
editor="Anderson, Clive W.
and Barnett, Vic
and Chatwin, Philip C.
and El-Shaarawi, Abdel H.",
title="Space and Space-Time Modeling using Process Convolutions",
booktitle="Quantitative Methods for Current Environmental Issues",
year="2002",
publisher="Springer London",
address="London",
pages="37--56",
isbn="978-1-4471-0657-9"
}

@article{Hakuba2024,
	Author = {Hakuba, Maria Z. and others},
	Doi = {10.1007/s10712-024-09849-5},
	Id = {Hakuba2024},
	Isbn = {1573-0956},
	Journal = {Surveys in Geophysics},
	Number = {6},
	Pages = {1721--1756},
	Title = {Trends and Variability in {Earth}'s Energy Imbalance and Ocean Heat Uptake Since 2005},
	Ty = {JOUR},
	Volume = {45},
	Year = {2024}
}

@article{Kuusela2018,
    author={Kuusela, Mikael and Stein, Michael L.},
    title={Locally stationary spatio-temporal interpolation of {A}rgo profiling float data},
    year={2018},
    journal={Proceedings of the Royal Society A},
    volume={474},
    number = {2220},
    pages = {20180400},
    doi={10.1098/rspa.2018.0400}
}

@article {LymanJohnson23,
      author = "John M. Lyman and Gregory C. Johnson",
      title = "Global High-Resolution Random Forest Regression Maps of Ocean Heat Content Anomalies Using In Situ and Satellite Data",
      journal = "Journal of Atmospheric and Oceanic Technology",
      year = "2023",
      publisher = "American Meteorological Society",
      address = "Boston MA, USA",
      volume = "40",
      number = "5",
      doi = "10.1175/JTECH-D-22-0058.1",
      pages=      "575-586"
}

@book{IPCC6,
   author = {IPCC},
   title = {Climate Change 2021 – The Physical Science Basis: Working Group I Contribution to the Sixth Assessment Report of the Intergovernmental Panel on Climate Change},
   publisher = {Cambridge University Press},
   DOI = {10.1017/9781009157896},
   year = {2023}
}

@article{RG2009,
title = {The 2004–2008 mean and annual cycle of temperature, salinity, and steric height in the global ocean from the {Argo} Program},
journal = {Progress in Oceanography},
volume = {82},
number = {2},
pages = {81-100},
year = {2009},
issn = {0079-6611},
doi = {10.1016/j.pocean.2009.03.004},
author = {Dean Roemmich and John Gilson}
}

@article{Kleiber2012,
title = {Nonstationary modeling for multivariate spatial processes},
journal = {Journal of Multivariate Analysis},
volume = {112},
pages = {76-91},
year = {2012},
issn = {0047-259X},
doi = {10.1016/j.jmva.2012.05.011},
author = {William Kleiber and Douglas Nychka}
}

@book{handbook, 
  title={Handbook of Spatial Statistics},
  author={Alan E. Gelfand and Montserrat Fuentes and Peter Guttorp},
  isbn={9781420072884},
  series={Chapman \& Hall/CRC Handbooks of Modern Statistical Methods},
  year={2010},
  publisher={CRC Press}
}

@ARTICLE{10.3389/fmars.2019.00432,
  
AUTHOR={Meyssignac, Benoit and others},   
	 
TITLE={Measuring Global Ocean Heat Content to Estimate the {Earth} Energy Imbalance},      
	
JOURNAL={Frontiers in Marine Science},      
	
VOLUME={6},           
pages={432},
YEAR={2019},      
	       
	
DOI={10.3389/fmars.2019.00432},      
	
ISSN={2296-7745}
}

@ARTICLE{Von_Schuckmann2016-rf,
  title     = "An imperative to monitor {Earth's Energy Imbalance}",
  author    = "von Schuckmann, K and Palmer, M D and Trenberth, K E and
               Cazenave, A and Chambers, D and Champollion, N and Hansen, J and
               Josey, S A and Loeb, N and Mathieu, P-P and Meyssignac, B and
               Wild, M",
  journal   = "Nat. Clim. Chang.",
  publisher = "Springer Science and Business Media LLC",
  volume    =  6,
  number    =  2,
  pages     = "138--144",
  month     =  feb,
  year      =  2016,
  language  = "en"
}

@ARTICLE{Johnson2022-sv,
  title     = "Global Oceans",
  author    = "Johnson, Gregory C and others",
  journal   = "Bull. Am. Meteorol. Soc.",
  publisher = "American Meteorological Society",
  volume    =  103,
  number    =  8,
  pages     = "S143--S192",
  month     =  aug,
  year      =  2022
}

@ARTICLE{Trenberth2018-jw,
  title     = "Hurricane {Harvey} links to ocean heat content and climate change
               adaptation",
  author    = "Trenberth, Kevin E and Cheng, Lijing and Jacobs, Peter and
               Zhang, Yongxin and Fasullo, John",
  journal   = "Earth's Future",
  publisher = "American Geophysical Union (AGU)",
  volume    =  6,
  number    =  5,
  pages     = "730--744",
  month     =  may,
  year      =  2018,
  copyright = "http://creativecommons.org/licenses/by-nc-nd/4.0/",
  language  = "en"
}

@article{Riser2016,
  title={Fifteen years of ocean observations with the global {Argo} array},
  author={Riser, Stephen C and Freeland, Howard J and Roemmich, Dean and Wijffels, Susan and Troisi, Ariel and Belb{\'e}och, Mathieu and Gilbert, Denis and Xu, Jianping and Pouliquen, Sylvie and Thresher, Ann and others},
  journal={Nature Climate Change},
  volume={6},
  number={2},
  pages={145--153},
  year={2016},
  publisher={Nature Publishing Group}
}

@INPROCEEDINGS{Brown2021-ji,
  title      = "Bridges-2: A platform for rapidly-evolving and data intensive
                research",
  booktitle  = "Practice and Experience in Advanced Research Computing",
  author     = "Brown, Shawn T and Buitrago, Paola and Hanna, Edward and
                Sanielevici, Sergiu and Scibek, Robin and Nystrom, Nicholas A",
  publisher  = "ACM",
  month      =  jul,
  year       =  2021,
  address    = "New York, NY, USA",
  conference = "PEARC '21: Practice and Experience in Advanced Research
                Computing",
  location   = "Boston MA USA"
}

@article {potentialenthalpy,
      author = "Trevor J. McDougall",
      title = "Potential Enthalpy: A Conservative Oceanic Variable for Evaluating Heat Content and Heat Fluxes",
      journal = "Journal of Physical Oceanography",
      year = "2003",
      publisher = "American Meteorological Society",
      address = "Boston MA, USA",
      volume = "33",
      number = "5",
      doi = {10.1175/1520-0485(2003)033<0945:PEACOV>2.0.CO;2},
      pages=      "945-963"
}

@article {Loeb2018,
      author = "Norman G. Loeb and David R. Doelling and Hailan Wang and Wenying Su and Cathy Nguyen and Joseph G. Corbett and Lusheng Liang and Cristian Mitrescu and Fred G. Rose and Seiji Kato",
      title = "Clouds and the {E}arth’s Radiant Energy System {(CERES)} Energy Balanced and Filled {(EBAF)} Top-of-Atmosphere {(TOA)} Edition-4.0 Data Product",
      journal = "Journal of Climate",
      year = "2018",
      publisher = "American Meteorological Society",
      address = "Boston MA, USA",
      volume = "31",
      number = "2",
      doi = "10.1175/JCLI-D-17-0208.1",
      pages=      "895-918"
}

@article {Kato2018,
      author = "Seiji Kato and Fred G. Rose and David A. Rutan and Tyler J. Thorsen and Norman G. Loeb and David R. Doelling and Xianglei Huang and William L. Smith and Wenying Su and Seung-Hee Ham",
      title = "Surface Irradiances of Edition 4.0 {Clouds and the Earth’s Radiant Energy System (CERES) Energy Balanced and Filled (EBAF)} Data Product",
      journal = "Journal of Climate",
      year = "2018",
      publisher = "American Meteorological Society",
      address = "Boston MA, USA",
      volume = "31",
      number = "11",
      doi = "10.1175/JCLI-D-17-0523.1",
      pages=      "4501-4527"
}

@incollection{Trenberth2020,
author = {Trenberth, Kevin E.},
publisher = {American Geophysical Union (AGU)},
isbn = {9781119548164},
title = {{ENSO} in the Global Climate System},
booktitle = {El Ni\~{n}o Southern Oscillation in a Changing Climate},
chapter = {2},
pages = {21-37},
doi = {https://doi.org/10.1002/9781119548164.ch2},
year = {2020}
}

@article{Yarger2022,
author = {Drew Yarger and Stilian Stoev and Tailen Hsing},
title = {{A functional-data approach to the Argo data}},
volume = {16},
journal = {The Annals of Applied Statistics},
number = {1},
publisher = {Institute of Mathematical Statistics},
pages = {216--246},
year = {2022},
doi = {10.1214/21-AOAS1477}
}

@misc{Salvana2022,
      title={{3D} Bivariate Spatial Modelling of {Argo} Ocean Temperature and Salinity Profiles}, 
      author={Mary Lai Salvana and Jian Cao and Mikyoung Jun},
      year={2022},
      howpublished={arXiv:2210.11611 [stat.AP]}, 
}

@article{Bolin2019,
   title={Multivariate Type {G Matérn} Stochastic Partial Differential Equation Random Fields},
   volume={82},
   ISSN={1467-9868},
   DOI={10.1111/rssb.12351},
   number={1},
   journal={Journal of the Royal Statistical Society Series B: Statistical Methodology},
   publisher={Oxford University Press (OUP)},
   author={Bolin, David and Wallin, Jonas},
   year={2019},
   month=dec, pages={215–239} }

@article {InsightsintoEarthsEnergyImbalancefromMultipleSources,
      author = "Kevin E. Trenberth and John T. Fasullo and Karina von Schuckmann and Lijing Cheng",
      title = "Insights into {E}arth's Energy Imbalance from Multiple Sources",
      journal = "Journal of Climate",
      year = "2016",
      publisher = "American Meteorological Society",
      address = "Boston MA, USA",
      volume = "29",
      number = "20",
      doi = "10.1175/JCLI-D-16-0339.1",
      pages= "7495-7505"
}

@misc{univariate,
    author = {Thea Sukianto and Mikael Kuusela and Donata Giglio and Anirban Mondal and Pulong Ma and Douglas W. Nychka},
    title = {Locally stationary {Argo} ocean heat content estimates: Modeling, validation and uncertainty quantification},
    howpublished = {arXiv:2606.31957 [stat.AP]},
    year = {2026}
}

@article{TEOS-10,
author = {McDougall, Trevor and Barker, P.M.},
year = {2011},
month = {01},
pages = {1-28},
title = {Getting started with {TEOS-10} and the {Gibbs Seawater (GSW) Oceanographic Toolbox}},
volume = {127},
journal = {SCOR/IAPSO WG}
}

@article{ONI,
author = {Anthony G. Bamston and Muthuvel Chelliah and Stanley B. Goldenberg},
title = {Documentation of a highly {ENSO}‐related {SST} region in the {Equatorial Pacific}: Research note},
journal = {Atmosphere-Ocean},
volume = {35},
number = {3},
pages = {367--383},
year = {1997},
publisher = {Taylor \& Francis},
doi = {10.1080/07055900.1997.9649597}
}

@article{Saduakhas2025,
author       = {Saduakhas, Damilya and Bolin, David and Jin, Xiaotian and
                  Simas, Alexandre B and Wallin, Jonas},
title = {Incorporating correlated nugget effects in multivariate spatial models: An application to {Argo} ocean data},
volume = {20},
journal = {The Annals of Applied Statistics},
number = {2},
publisher = {Institute of Mathematical Statistics},
pages = {1187--1207},
year = {2026},
doi = {10.1214/26-AOAS2194}
}

@software{matlab-optim,
year = {2021},
author = {{The MathWorks Inc.}},
title = {Optimization Toolbox version: 9.1 (R2021a)},
publisher = {The MathWorks Inc.},
address = {Natick, Massachusetts, United States}
}

@ARTICLE{Wong2020,
  
AUTHOR={Wong, Annie P. and others},   
TITLE={{Argo} Data 1999–2019: Two Million Temperature-Salinity Profiles and Subsurface Velocity Observations From a Global Array of Profiling Floats},
        
JOURNAL={Frontiers in Marine Science},
        
VOLUME={7},
pages={700},
YEAR={2020},
DOI={10.3389/fmars.2020.00700},

ISSN={2296-7745}}

@article{Roemmich2009_ArgoProgram,
 ISSN = {10428275, 2377617X},
 author = {Dean Roemmich and Gregory C. Johnson and Stephen Riser and Russ Davis and John Gilson and W. Brechner Owens and Silvia L. Garzoli and Claudia Schmid and Mark Ignaszewski},
 journal = {Oceanography},
 number = {2},
 pages = {34--43},
 publisher = {Oceanography Society},
 title = {The {Argo} Program: Observing the Global Ocean with Profiling Floats},
 urldate = {2022-12-09},
 volume = {22},
 year = {2009}
}

@article{ArgoDataJan2023,
   author={Argo},
   title={{Argo} float data and metadata from {Global Data Assembly Centre (Argo GDAC)} - {S}napshot of {Argo} {GDAC} of {J}anuary 10th 2023},
   year={2023},
   journal={SEANOE},
   doi={10.17882/42182#98916},
}

@article{Korte-Stapff2025,
    author = {Korte-Stapff, Moritz and Yarger, Drew and Stoev, Stilian and Hsing, Tailen},
    title = {A functional regression model for heterogeneous {BioGeoChemical Argo data in the Southern Ocean}},
    journal = {Journal of the Royal Statistical Society Series C: Applied Statistics},
    volume = {75},
    number = {1},
    pages = {79-99},
    year = {2025},
    month = {06},
    issn = {0035-9254},
    doi = {10.1093/jrsssc/qlaf036}
}

@Article{von_Schuckmann2023,
AUTHOR = {von Schuckmann, K. and others},
TITLE = {Heat stored in the {E}arth system 1960--2020: where does the energy go?},
JOURNAL = {Earth System Science Data},
VOLUME = {15},
YEAR = {2023},
NUMBER = {4},
PAGES = {1675--1709},
DOI = {10.5194/essd-15-1675-2023}
}

@article{ZammitMangion2025,
   author = "Zammit-Mangion, Andrew and Sainsbury-Dale, Matthew and Huser, Raphaël",
   title = "Neural Methods for Amortized Inference", 
   journal= "Annual Review of Statistics and Its Application",
   year = "2025",
   volume = "12",
   pages = "311-335",
   doi = "https://doi.org/10.1146/annurev-statistics-112723-034123",
   publisher = "Annual Reviews",
}

\end{document}